\documentclass[11pt,a4paper]{article}
\usepackage[T1]{fontenc}
\usepackage{lmodern}
\usepackage[utf8]{inputenc}
\usepackage{amsmath,amssymb,amsthm}
\usepackage[a4paper,textwidth=126mm,textheight=195mm,centering]{geometry}
\usepackage{graphicx}
\usepackage{booktabs}
\usepackage{array}
\usepackage{float}
\usepackage[section]{placeins}
\usepackage{tikz}
\usetikzlibrary{arrows.meta,patterns,calc}
\usepackage[hidelinks]{hyperref}
\usepackage{caption}
\theoremstyle{plain}
\newtheorem{theorem}{Theorem}
\newtheorem{proposition}[theorem]{Proposition}
\newtheorem{lemma}[theorem]{Lemma}

\theoremstyle{definition}

\newcommand{\alttext}[1]{\par\smallskip{\footnotesize\setlength{\leftskip}{0pt}%
  \setlength{\rightskip}{0pt}\setlength{\parfillskip}{0pt plus 1fil}%
  \noindent Alt text: #1\par}}
\newcommand{\ZZ}{\mathbb{Z}}
\newcommand{\FF}{\mathbb{F}_2}
\newcommand{\Gmax}[1]{\Gamma^{#1}_{\max}}

\newcommand{\Ak}[1]{A^{\leq #1}_1}

\title{Persistent $A$-homotopy of networks: free groups and periodic structures}
\author{Milan Rajkovi\'{c}\\[0.5em]
	{\small Institute of Nuclear Sciences Vin\v{c}a, University of Belgrade,}\\
	{\small 11001 Belgrade, Serbia}\\[0.5em]
	{\small e-mail: \href{mailto:milanqr@proton.me}{milanqr@proton.me}}}
\date{}

\begin{document}
\maketitle

\begin{abstract}
We study how loop classes in a network change when cycles up to a
prescribed length $k$ are declared contractible. The resulting sequence of
quotient groups contains the fundamental group of the clique complex at
$k=3$ and the graph $A$-fundamental group at $k=4$. Its mod-two
abelianisation is first homology persistence of a cell filtration, and its
barcode has the lengths in a minimum cycle basis as death scales. We ask when this barcode determines the groups. If
every group is free, it determines the groups and the maps between them.
We give a test that certifies freeness by eliminating generators and proves
non-freeness through torsion, homomorphism counts and the Hopfian property.
Every group was certified free for nine benchmark networks, for
Watts--Strogatz, scale-free, clustered and modular models, for a random
geometric graph in the unit square and for street networks. We prove that
sparse Erd\H{o}s--R\'enyi graphs have free groups at each fixed length with
probability tending to one, and that dominated vertices, common in
clustered networks, can be deleted without changing the groups. Groups
that are not free appear when short cycles close up into a surface. In
random geometric graphs on a torus the group is $\ZZ^2$ over a window of
lengths, and in six of nine samples it changes from $F_2$ to $\ZZ^2$ at a
length where no bar ends. A network repeated over $T$ periodic time steps
has the group of the network times $\ZZ$ at lengths $4\le k<T$, whereas
attaching a separate cycle of length $T$ gives the same barcode and free
groups. Where the groups are free, we use the barcode to analyse
Watts--Strogatz networks, the benchmarks and loops of cliques around hubs.
\end{abstract}


\section{Introduction}
\label{sec:intro}

A network records pairwise relations, but many questions concern how these
relations fit together. Simplicial complexes provide one way to represent
higher-order organisation~\cite{Battiston2020}. In the clique complex of a
graph, every complete subgraph is filled by a simplex. Persistent homology
then follows homology classes as a threshold changes the
network~\cite{HMR2009,Petri2013}. It distinguishes classes that disappear
quickly from those that survive over a substantial threshold range.

There is a second modelling choice. Even on a fixed network, one can ask
which rearrangements of a loop should be allowed. A chordless four-cycle
and a chordless thirty-cycle each contribute one first-homology class to
their clique complexes. If exchanging the two routes across a square is
regarded as a local rearrangement, the four-cycle becomes contractible,
whereas the isolated thirty-cycle does not. The distinction comes from the
allowed moves. It need not be a property of physical dynamics unless those
moves have an interpretation in the system under study.

Discrete $A$-homotopy supplies a precise language for this choice.
Following Atkin's work on the connectivity of simplicial
complexes~\cite{Atkin1974,Atkin1977}, Barcelo, Kramer, Laubenbacher and Weaver
developed $A$-theory for complexes~\cite{BKLW2001,BL2005}. A related theory
is defined for graphs~\cite{BBdLL2006}. At the level of based loops, triangle
shortcuts and square slides generate the non-degenerate local moves. The
resulting group retains the order in which loops are traversed, information
that first homology discards through abelianisation.
Section~\ref{sec:primer} gives the definitions used here, starting from
ordinary graphs. The companion paper~\cite{Rajkovic2026a} gives a fuller
introduction to $A$-homotopy through worked examples.

Persistence has already been combined with discrete homotopy and discrete
homology along a distance scale. M\'emoli and Zhou study persistent homotopy
groups of compact metric spaces~\cite{MemoliZhou}, and Kapulkin and Kershaw
propose persistent discrete homology for finite metric and semi-metric
data~\cite{KapulkinKershaw}. Both works start from distances, not from a
given network or simplicial complex. M\'emoli and Zhou compare several
constructions of the persistent fundamental group, one through
Vietoris--Rips complexes of the metric space and one through discrete
fundamental groups. Graphs appear in their examples with the shortest-path
distance. Kapulkin and Kershaw join two data points when their distance is
at most the scale and compute the discrete homology of the resulting
graphs, and Vietoris--Rips complexes serve them as a comparison. In both, the
scale changes the space or graph under study. Here the network is held fixed, and we let
the maximum length of an allowed cycle relation vary instead. At scale
$k$, all simple cycles of length at most $k$ are declared contractible.
We call the resulting sequence of groups \emph{persistent $A$-homotopy in
coordination scale}. Its algebraic theory, including the interval
decomposition, the minimum-cycle-basis description and a stability bound for
edge insertions, is developed in the companion paper~\cite{Rajkovic2026b},
where $k$ is called the relation-length scale. The present paper applies the
construction to networks. The familiar graph $A$-fundamental group occurs at
$k=4$. The term refers in this paper to fundamental groups and their maps.
We do not construct a persistence theory for higher $A$-groups. The scale
$k$ counts edges and specifies the permitted relations. It is not an
elapsed time, an interaction strength or a measured coordination capacity.

For computation one uses the mod-two abelianisation of this sequence. Its
barcode is determined by a minimum cycle basis. The mod-two summary is also
ordinary first homology persistence of a cell filtration obtained by
attaching the allowed cycle fillings. Persistent $A$-homotopy therefore has a direct connection to
persistent homology. The barcode is a linear summary, however, and it
forgets the order in which loops are composed. The question that organises
this paper is whether the barcode determines the groups, and what the
groups record when it does not.

\begin{figure}[!t]
\centering
\includegraphics[width=\textwidth]{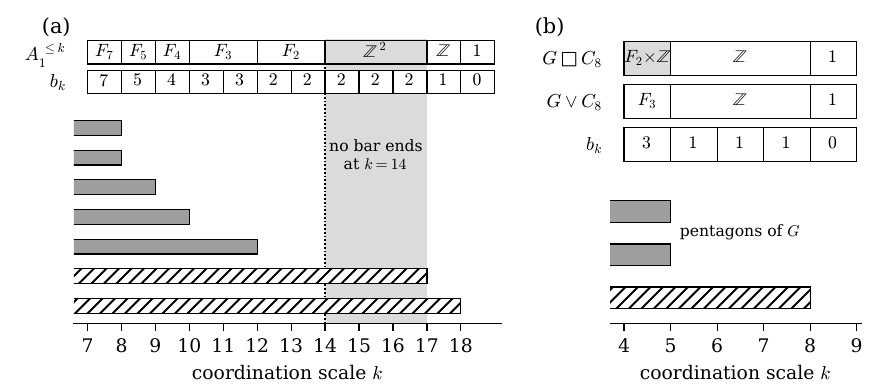}
\caption{\textbf{Two cases in which the barcode does not determine the groups.} (a) A random geometric graph on the torus with $n=800$ points and radius $r=0.07$, the first row of Table~\ref{tab:torus}. Shown are the seven bars alive at $k=7$, all of which begin at $k=2$. The rows above the bars give the group $\Ak{k}(G)$ and the dimension $b_k$. At $k=14$ the group changes from $F_2$ to $\ZZ^2$, but no bar ends there. The hatched bars survive through the window $14\le k\le16$. (b) The Florentine marriage network $G$ over eight periodic time steps, $G\mathbin{\square}C_8$, and the wedge $G\vee C_8$ obtained by attaching a separate cycle of length eight at one vertex. For $k\ge4$ the two graphs have the same barcode, shown here. The grey bars can be represented by the two pentagons of $G$ and the hatched bar by a loop around the period. Grey cells mark groups that are not free. The groups in (a) are read from the reduced presentations of the test of Section~\ref{sec:freetest}. In (b) the test certifies the free groups and shows that the product group at $k=4$ is not free, and Proposition~\ref{prop:product} identifies it as $F_2\times\ZZ$.}
\label{fig:one}
\alttext{Two horizontal barcodes against coordination scale $k$, each with rows of labelled cells above the bars. Panel (a) runs from $k=7$ to $18$. Seven bars enter from the left. Five grey bars end at $8$, $8$, $9$, $10$ and $12$, and two hatched bars end at $17$ and $18$. The row labelled $b_k$ reads $7,5,4,3,3,2,2,2,2,2,1,0$ for $k=7$ to $18$. The group row reads $F_7$, $F_5$ and $F_4$ for $k=7$, $8$ and $9$, $F_3$ for $10$ and $11$, $F_2$ for $12$ and $13$, $\mathbb{Z}^2$ for $14$ to $16$ in a grey cell, $\mathbb{Z}$ for $17$ and $1$ for $18$. A grey vertical band covers $k=14$ to $17$, with a dotted line at $k=14$ and the note that no bar ends at $k=14$. Panel (b) runs from $k=4$ to $8$. Two grey bars, labelled as pentagons of $G$, end at $5$, and a hatched bar ends at $8$. The rows read $F_2\times\mathbb{Z}$ in a grey cell, then $\mathbb{Z}$ for $k=5$ to $7$ and $1$ at $8$ for $G\,\square\,C_8$, the same with $F_3$ at $k=4$ for $G\vee C_8$, and $b_k=3,1,1,1,0$.}
\end{figure}

The answer depends on whether the groups are free. A free group is
determined by its rank, so if every group in the sequence is free, the
barcode determines the groups and, by Nielsen's reduction method, the maps
between them (Proposition~\ref{prop:freetower}). Section~\ref{sec:free}
gives a test that certifies freeness by eliminating generators from a
presentation and proves non-freeness through torsion, homomorphism counts
into the symmetric group $S_3$ and the Hopfian property. For every graph in
Table~\ref{tab:free}, every group in the sequence is free. These graphs
include nine benchmark networks with their clique-connectivity graphs and
randomisations, Watts--Strogatz networks, Barab\'asi--Albert, Holme--Kim,
configuration-model and hyperbolic graphs with heavy-tailed degrees, LFR
and stochastic block models, duplication--divergence graphs, a random
geometric graph in the unit square and street networks. Two results
account for much of this. In sparse Erd\H{o}s--R\'enyi graphs the short
cycles are vertex-disjoint with probability tending to one, which makes the
group at each fixed length free (Proposition~\ref{prop:sparse}). In
clustered networks, a vertex whose neighbourhood lies in that of a
neighbour can be deleted without changing the groups
(Lemma~\ref{lem:dominated}). Freeness is nevertheless not a general
property of random graphs. Two statements of Babson~\cite{Babson2012}
together imply that, for Erd\H{o}s--R\'enyi graphs with edge probability
roughly between $n^{-1/2}$ and $n^{-1/3}$, the group at length three is
non-trivial while $b_3=0$, with probability tending to one. Such a group is
not free (Section~\ref{sec:whyfree}).

Section~\ref{sec:periodic} constructs networks whose groups are not free.
In each of them, short cycles close up into a torus through two
independent periodic directions. Filling the plaquettes of a periodic
square grid gives the torus group $\ZZ^2$. In random geometric graphs on a
torus the group is $\ZZ^2$ over a window of lengths, and in six of nine
samples it changes from $F_2$ to $\ZZ^2$ at a length at which no bar ends
(Fig.~\ref{fig:one}a). Random shortcuts added to a periodic grid contribute
free generators and give $\ZZ^2\ast F_s$ when no shortcut lies on a
triangle or a square. A network $G$ observed over a
periodic schedule of $T$ time steps is the Cartesian product
$G\mathbin{\square}C_T$. Its group at length $k$ is $\Ak{k}(G)\times\ZZ$
for $4\le k<T$, whereas attaching a separate cycle of length $T$ to $G$
gives the same barcode and a free group (Fig.~\ref{fig:one}b). In these
examples the group records commutation between loops, which the barcode
cannot see.

Where the groups are free, the barcode describes the whole construction,
and Section~\ref{sec:complex} analyses it for network models and data. In
Watts--Strogatz networks, the largest death scale and the mean path length,
both normalised by their values for the ring lattice, decrease with
rewiring, and the first lies below the second over the sampled range.
Benchmark networks differ in how much clique homology the square relations
remove and in how far the remaining bars extend. Around a hub, the
connectivity graph of maximal cliques can carry loops even though the ego
clique complex is contractible.

Section~\ref{sec:primer} explains $A$-homotopy through loop moves on
ordinary graphs and then extends it to networks of groups.
Section~\ref{sec:persistent} introduces the filtration, its computable
summary and the condition under which the summary determines the groups.
Section~\ref{sec:free} tests that condition, Section~\ref{sec:periodic}
constructs the periodic examples and Section~\ref{sec:complex} analyses the
barcodes of network models and data. The discussion considers where the
group-valued construction adds information to network analysis and what
would be needed to attach a dynamical interpretation to it.

\section{A primer on \texorpdfstring{$A$}{A}-homotopy}
\label{sec:primer}

This section introduces the ideas used in the rest of the paper,
starting from an ordinary network. Section~\ref{sec:moves} describes
loops in a graph and the local moves that deform them.
Section~\ref{sec:certificates} turns these moves into a calculation and
gives simple tests for deciding whether a loop can be removed.
Section~\ref{sec:groups} applies the same moves to networks whose
vertices are groups of nodes, such as cliques or the participants of
an event.

\subsection{Loops and local moves}
\label{sec:moves}

All graphs in the mathematical statements are finite, simple and
undirected. The cycle rank
\[
 \beta_1(G)=|E|-|V|+c(G),
\]
where $c(G)$ is the number of connected components, counts the
independent cycles of $G$. Homology uses the field $\FF$ unless $\ZZ$
is displayed explicitly. A group $F_r$ denotes a free group on $r$
generators, not a field. Fundamental groups are based in a specified
component. When a numerical rank is given for a disconnected graph, it
is the sum of the ranks over components.

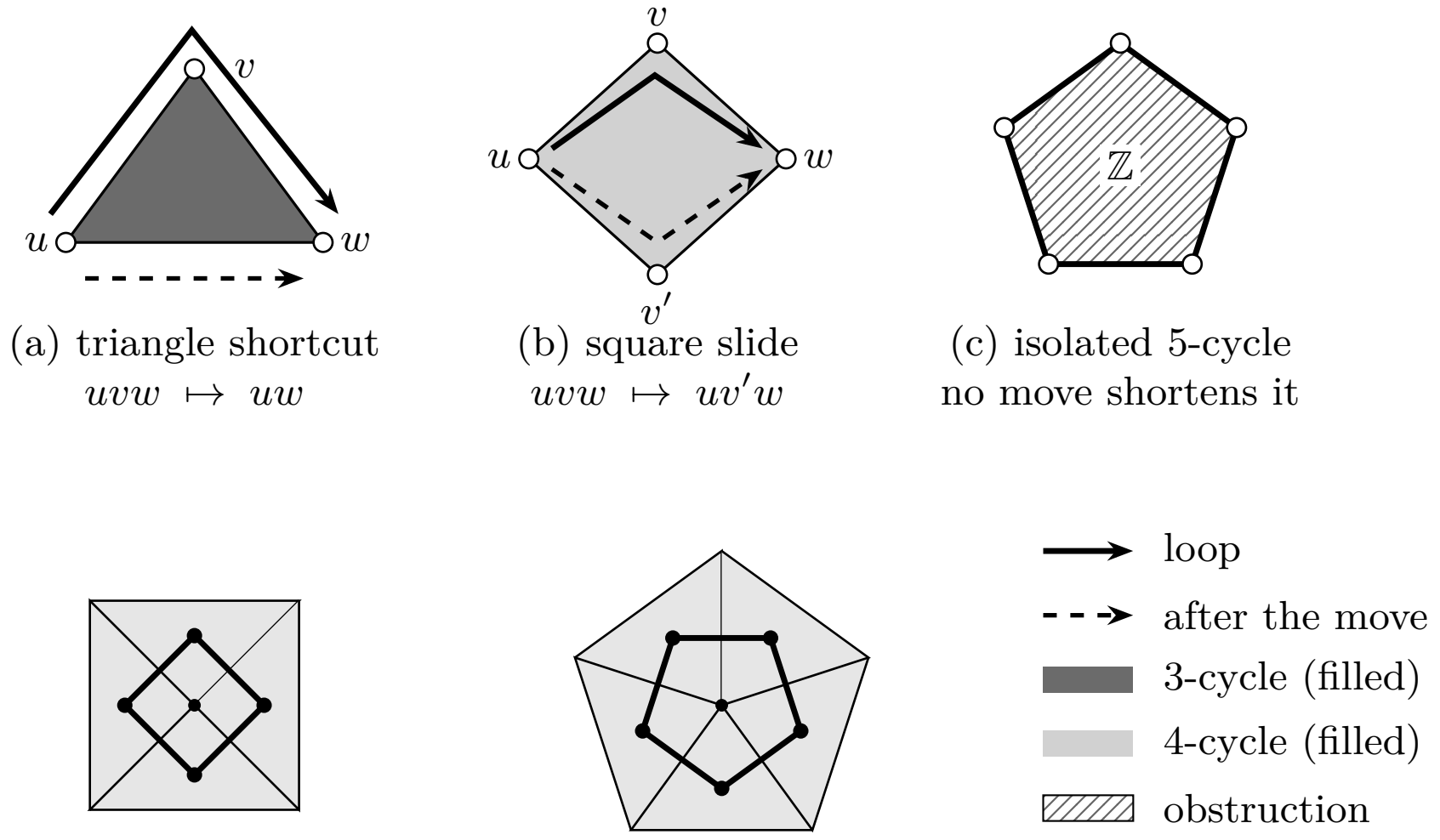
\begin{figure}[t]
\centering
\begin{tikzpicture}[x=1cm,y=1cm, every node/.style={font=\footnotesize},
  vx/.style={circle, draw, fill=white, inner sep=0pt, minimum size=4.2pt, line width=0.5pt},
  loop/.style={line width=1.3pt, -{Stealth[length=2mm]}},
  moved/.style={line width=1.0pt, dashed, -{Stealth[length=2mm]}}]
\begin{scope}[shift={(0,0)}]
  \coordinate (s) at (0,0); \coordinate (t) at (1,1.35); \coordinate (r) at (2,0);
  \fill[black!58] (s) -- (t) -- (r) -- cycle;
  \draw[loop] ($(s)+(-0.12,0.22)$) -- ($(t)+(-0.02,0.3)$) -- ($(r)+(0.12,0.22)$);
  \draw[line width=0.6pt] (s) -- (t) -- (r) -- (s);
  \draw[moved] ($(s)+(0.15,-0.28)$) -- ($(r)+(-0.15,-0.28)$);
  \node[vx] at (s) {}; \node[vx] at (t) {}; \node[vx] at (r) {};
  \node[left] at (s) {$u$}; \node[right=5pt] at (t) {$v$}; \node[right] at (r) {$w$};
  \node[align=center] at (1,-0.95) {(a) triangle shortcut\\ $uvw\;\mapsto\;uw$};
\end{scope}
\begin{scope}[shift={(3.6,0)}]
  \coordinate (s) at (0,0.65); \coordinate (t) at (1,1.55); \coordinate (r) at (2,0.65); \coordinate (u) at (1,-0.25);
  \fill[black!18] (s) -- (t) -- (r) -- (u) -- cycle;
  \draw[line width=0.6pt] (s) -- (t) -- (r) -- (u) -- cycle;
  \draw[loop] ($(s)+(0.18,0.08)$) -- ($(t)+(-0.02,-0.25)$) -- ($(r)+(-0.18,0.08)$);
  \draw[moved] ($(s)+(0.18,-0.08)$) -- ($(u)+(0.0,0.25)$) -- ($(r)+(-0.18,-0.08)$);
  \node[vx] at (s) {}; \node[vx] at (t) {}; \node[vx] at (r) {}; \node[vx] at (u) {};
  \node[left] at (s) {$u$}; \node[above] at (t) {$v$}; \node[right] at (r) {$w$}; \node[below] at (u) {$v'$};
  \node[align=center] at (1,-0.95) {(b) square slide\\ $uvw\;\mapsto\;uv'w$};
\end{scope}
\begin{scope}[shift={(8.2,0.6)}]
  \foreach \i in {0,...,4} { \coordinate (p\i) at ({90+72*\i}:0.95); }
  \fill[pattern=north east lines, pattern color=black!60] (p0) -- (p1) -- (p2) -- (p3) -- (p4) -- cycle;
  \draw[line width=1.3pt] (p0) -- (p1) -- (p2) -- (p3) -- (p4) -- cycle;
  \foreach \i in {0,...,4} { \node[vx] at (p\i) {}; }
  \node[fill=white, inner sep=1.5pt] at (0,0) {$\mathbb{Z}$};
  \node[align=center] at (0,-1.55) {(c) isolated $5$-cycle\\ no move shortens it};
\end{scope}
\begin{scope}[shift={(1.0,-3.6)}]
  \foreach \i in {0,...,3} { \coordinate (w\i) at ({45+90*\i}:1.15); }
  \foreach \i/\j in {0/1,1/2,2/3,3/0} { \fill[black!10] (0,0) -- (w\i) -- (w\j) -- cycle; \draw[line width=0.5pt] (0,0) -- (w\i) -- (w\j); }
  \foreach \i/\j in {0/1,1/2,2/3,3/0} { \coordinate (c\i) at ($0.333*(w\i)+0.333*(w\j)$); }
  \draw[line width=1.3pt] (c0) -- (c1) -- (c2) -- (c3) -- cycle;
  \foreach \i in {0,...,3} { \fill (c\i) circle (1.7pt); }
  \fill (0,0) circle (1.4pt);
  \node[align=center] at (0,-1.65) {(d) fan $F_4$, $\Gamma^1_{\max}=C_4$\\ $\widetilde H_*(|F_4|)=0$, \ $A^1_1=1$};
\end{scope}
\begin{scope}[shift={(5.1,-3.6)}]
  \foreach \i in {0,...,4} { \coordinate (w\i) at ({90+72*\i}:1.2); }
  \foreach \i/\j in {0/1,1/2,2/3,3/4,4/0} { \fill[black!10] (0,0) -- (w\i) -- (w\j) -- cycle; \draw[line width=0.5pt] (0,0) -- (w\i) -- (w\j); }
  \foreach \i/\j in {0/1,1/2,2/3,3/4,4/0} { \coordinate (c\i) at ($0.333*(w\i)+0.333*(w\j)$); }
  \draw[line width=1.3pt] (c0) -- (c1) -- (c2) -- (c3) -- (c4) -- cycle;
  \foreach \i in {0,...,4} { \fill (c\i) circle (1.7pt); }
  \fill (0,0) circle (1.4pt);
  \node[align=center] at (0,-1.65) {(e) fan $F_5$, $\Gamma^1_{\max}=C_5$\\ $\widetilde H_*(|F_5|)=0$, \ $A^1_1\cong\mathbb{Z}$};
\end{scope}
\begin{scope}[shift={(7.6,-3.0)}]
  \draw[loop] (0,0.6) -- (0.7,0.6); \node[right] at (0.8,0.6) {loop};
  \draw[moved] (0,0.1) -- (0.7,0.1); \node[right] at (0.8,0.1) {after the move};
  \fill[black!58] (0,-0.5) rectangle (0.7,-0.3); \node[right] at (0.8,-0.4) {$3$-cycle (filled)};
  \fill[black!18] (0,-1.0) rectangle (0.7,-0.8); \node[right] at (0.8,-0.9) {$4$-cycle (filled)};
  \fill[pattern=north east lines, pattern color=black!60] (0,-1.5) rectangle (0.7,-1.3); \draw (0,-1.5) rectangle (0.7,-1.3);
  \node[right] at (0.8,-1.4) {obstruction};
\end{scope}
\end{tikzpicture}
\caption{\textbf{Loop moves and fan examples.} (a, b) The triangle shortcut and the square slide. Together with stutters and backtracks they generate $A$-homotopy of based loops. (c) For the isolated graph $C_5$, the loop generates $A_1\cong\ZZ$. The claim does not apply to every pentagon inside a larger graph. (d, e) The fans $F_4$ and $F_5$ of Section~\ref{sec:groups} are cones on $C_4$ and $C_5$, with reduced homology zero. Their connectivity graphs of maximal triangles are the black cycles. The first has trivial $A$-group and the second has group $\ZZ$. The hatching marks a surviving loop, not a filled cell.}
\label{fig:primer}
\alttext{Panel (a) shows a dark grey triangle with vertices $u$, $v$ and $w$; a heavy arrow runs from $u$ through $v$ to $w$, and a dashed arrow runs directly from $u$ to $w$. Panel (b) shows a light grey square with vertices $u$, $v$, $w$ and $v'$; a heavy arrow runs through $u$, $v$ and $w$, and a dashed arrow runs through $u$, $v'$ and $w$. Panel (c) shows a hatched pentagon labelled $\mathbb{Z}$. Panels (d) and (e) show fans of four and five shaded triangles around a central hub, with a heavy cycle joining the centres of neighbouring triangles: a square in (d) and a pentagon in (e). A legend identifies the loop, the loop after the move, filled triangles, filled squares and the hatching for an obstruction.}
\end{figure}

A loop in $G$ based at a vertex $v_0$ is a closed walk that starts and
ends at $v_0$. Vertices and edges may be visited more than once. One
can picture a route that a message or a piece of business follows
through the network before it returns to its starting point. Two loops
are $A$-homotopic when one can be turned into the other by a finite
sequence of the following four moves, each of which may be applied in
either direction:
\begin{itemize}\raggedright
\item \emph{stutter}: repeat a vertex, or delete an immediate
repetition, $(\ldots,u,u,\ldots)\leftrightarrow(\ldots,u,\ldots)$.
\item \emph{backtrack}: insert or delete a step to a neighbour and
straight back, $(\ldots,u,v,u,\ldots)\leftrightarrow(\ldots,u,\ldots)$.
\item \emph{triangle shortcut}: if $u$, $v$ and $w$ are pairwise
adjacent, $(\ldots,u,v,w,\ldots)\leftrightarrow(\ldots,u,w,\ldots)$.
\item \emph{square slide}: if $(u,v,w,v',u)$ is a four-cycle,
$(\ldots,u,v,w,\ldots)\leftrightarrow(\ldots,u,v',w,\ldots)$.
\end{itemize}
Figure~\ref{fig:primer}(a,~b) shows the last two moves. Stutters and
backtracks change only how long a walk pauses and whether it makes a
detour that it immediately undoes. These two moves generate the ordinary
homotopy of edge paths, and the resulting classes of loops form the
fundamental group $\pi_1(G,v_0)$. The triangle shortcut and the square
slide are what $A$-homotopy adds: a detour around a triangle or a square
is treated as a local rearrangement that can be removed. The formal
definition~\cite{BKLW2001,BBdLL2006} uses a rectangular grid of vertices
whose first and last rows are the two loops. Every row is a loop at
$v_0$, shorter rows are padded by repetitions, and horizontally or
vertically neighbouring entries must be equal or adjacent. The four
moves generate the same equivalence.

Following one loop by another defines a product of loops. With this
product the classes of loops form a group, the graph $A$-fundamental
group $A_1(G,v_0)$. Its identity element is the class of the loop that
stays at $v_0$. A loop is trivial when it is $A$-homotopic to that
constant loop.

The cycle graphs give the first examples. Consider the graph $C_n$ on
its own. The loop once around $C_4$ is the boundary of a square, and a
square slide followed by backtracks reduces it to the constant loop,
so $A_1(C_4)=1$. For $n\ge5$ the graph has no triangles or squares,
only stutters and backtracks are available, and these do not change the
number of times a loop winds around the cycle. Hence $A_1(C_n)\cong\ZZ$
for $n\ge5$, generated by one turn around the cycle
(Fig.~\ref{fig:primer}c). This conclusion need not hold for a cycle
embedded in a larger graph, where other vertices may supply a
contraction.

Persistent homology of networks detects loops through first homology.
The two approaches differ in what they fill and in what they record.
The first homology of the clique complex, defined in
Section~\ref{sec:groups}, treats every triangle as filled and leaves
every chordless square open, whereas $A$-homotopy fills squares as
well. First homology records how many times a loop runs around each
independent cycle, in whatever order. The group $A_1(G,v_0)$ also keeps
the order in which the cycles are traversed. If $a$ and $b$ are loops at
$v_0$ around two different cycles, the loop $aba^{-1}b^{-1}$ runs around
$a$, then $b$, then $a$ in reverse and finally $b$ in reverse. It is zero
in homology, but in the group it is trivial exactly when $a$ and $b$
commute. For a connected graph the two descriptions are linked. Fill
every triangle and square of $G$ to obtain the space $Y_4(G)$ of
Section~\ref{sec:certificates}. Its first homology is the
abelianisation of $A_1(G,v_0)$, that is, the group with the order of
traversal forgotten. Section~\ref{sec:persistent} uses the mod-two form
of this statement, Eq.~\eqref{eq:linearisation}. In the Florentine
marriage network of Fig.~\ref{fig:florentine}, the $A$-group is free on
two generators. These do not commute, so the order of traversal carries
information there that homology does not record.

\subsection{Filling triangles and squares}
\label{sec:certificates}

Deciding by hand whether two loops are related by moves is impractical
in a large network. A description of the group in terms of filled cells
reduces the question to a calculation. Let $Y_4(G)$ be obtained from
$G$ by attaching a disc along every simple triangle and quadrilateral,
as if a membrane were stretched over each of them. Only the component
containing the basepoint is used for a fundamental group.

\begin{theorem}[Barcelo, Kramer, Laubenbacher and Weaver~{\cite[Prop.~5.12]{BKLW2001}}]
\label{thm:BL}
For a graph $G$ and a vertex $v_0$,
\[
 A_1(G,v_0)\cong\pi_1(Y_4(G),v_0)
 \cong\pi_1(G,v_0)/N_{\le4},
\]
where $N_{\le4}$ is the normal closure of the triangle and square loops.
\end{theorem}

Theorem~\ref{thm:BL} restates, for an arbitrary graph, the result that
Barcelo, Kramer, Laubenbacher and Weaver prove for the connectivity
graphs of simplicial complexes~\cite[Prop.~5.12]{BKLW2001}. See also the
overview by Barcelo and Laubenbacher~\cite{BL2005} and the formulation
using maximal simplices in~\cite[Sec.~2]{BS2008}. The second isomorphism
is the standard presentation of the fundamental group of a
two-dimensional cell complex.

In words, a loop is trivial in $A_1(G)$ exactly when it can be pulled
to a point in $Y_4(G)$, across the membranes. In practice, one
starts from the fundamental group of a connected graph, which is free on
$\beta_1(G)$ generators, and imposes one relation for each triangle and
each square. To write a cycle as a relation, a path from
$v_0$ to the cycle turns it into a based loop. Changing that path
conjugates the relation and leaves its normal closure unchanged. We
recall why the cell description holds: the elementary moves insert or
remove precisely the boundaries of the attached discs, in addition to
backtracks and stutters.

A nonzero element of $H_1(Y_4(G);\FF)$ proves that $A_1(G)$ is
non-trivial. Zero homology does not prove that the group is trivial.
The following sufficient tests make stronger conclusions possible.

\begin{lemma}[Certificates]
\label{lem:cert}
Let $G$ be a connected graph.
\begin{enumerate}
\item[(a)] If an edge $e$ belongs to no triangle or square and a closed
walk traverses it an odd number of times, that walk is non-trivial in
$A_1(G)$.
\item[(b)] Mark a spanning tree. Repeatedly mark the sole unmarked edge
of a triangle or square whose other edges are marked. If every edge
becomes marked, then $A_1(G)=1$.
\item[(c)] Attach all triangles and chordless squares. If every such
cell can be removed, together with an edge on no other remaining cell,
the group is free of rank $\beta_1(G)$ minus the number of cells removed.
\end{enumerate}
\end{lemma}

\begin{proof}
For (a), the cochain equal to one on $e$ and zero on other edges vanishes
on all cell boundaries but evaluates to one on the walk. For (b), collapse
the spanning tree in the group presentation. Every new marking uses a
relator to show that its last unmarked generator is trivial. For (c),
chorded square relations already follow from triangle relations, so they
may be omitted without changing the group. Each specified removal is a
free-edge collapse. It preserves homotopy type and decreases the graph
cycle rank by one. If all cells are removed, the remaining graph has the
stated free fundamental group.
\end{proof}

The three tests have simple readings. In (a), the edge $e$ lies on no
triangle or square, so the only move that changes the number of times a
loop crosses $e$ is a backtrack, which changes it by two. The parity of
the number of crossings is therefore fixed, and a loop that crosses $e$
an odd number of times cannot be trivial. We call this an edge cochain
certificate. In (b), the edges of the spanning tree carry no generators.
A triangle or square relation in which every generator but one is
already known to be trivial forces the last one to be trivial as well.
We call this spanning-forest propagation. In (c), each removal is a
free-edge collapse, which peels off a cell through an edge that no other
cell uses. For disconnected graphs these tests apply componentwise. A failed
test in (b) or (c) is inconclusive. In particular, a positive mod-two
rank and a successful collapse justify different amounts of information
about a group.

\subsection{Networks of groups}
\label{sec:groups}

Many networks are built from groups: committees, co-authored papers,
events with their participants, or the cliques of a graph. A simplicial complex $\Delta$ records such data. It is a family
of finite vertex sets closed under taking subsets, and a set in the
family is a simplex. A simplex with $q+1$ vertices has dimension $q$.
The clique complex $X(G)$ of a graph consists of all cliques of $G$.
Its triangles are filled faces, and a clique on four vertices is a
tetrahedron. A bipartite relation gives another construction. For
example, each event can supply the simplex of people who attended it,
together with all its faces. This is one of the two Dowker complexes of
the relation.

To study how the groups themselves are connected, one builds a graph
whose vertices are simplices. Two simplices are $q$-near when they share
at least $q+1$ vertices, and $q$-connected when a chain of such near
pairs joins them. Let $\Gamma^q(\Delta)$ have the simplices of dimension
at least $q$, which we call eligible, as vertices, with edges joining
distinct $q$-near pairs.
Its subgraph on the maximal simplices of dimension at least $q$ is the
connectivity graph $\Gmax{q}(\Delta)$. For committees, $q=1$ joins two
committees that share at least two members. Atkin's $Q$-vector records
the number of components of these graphs. In the language of
$A$-homotopy, the components are the zeroth $A$-homotopy classes
$A^q_0(\Delta)$, and
\[
 Q_q(\Delta)=|A^q_0(\Delta)|=c(\Gamma^q(\Delta)).
\]
The count is zero when there are no eligible simplices.

For the Southern Women data~\cite{Davis1941,NXdata}, take the fourteen events as
simplices on the eighteen women. The resulting counts for
$q=0,1,\ldots,13$ are
\[
 (1,1,1,2,1,2,1,3,2,3,2,2,1,1).
\]
These counts say how many pieces remain connected at each overlap level.
The next question is whether loops within a piece can be contracted.

Loops of simplices are treated exactly as loops of vertices in
Section~\ref{sec:moves}, with adjacency replaced by $q$-nearness. For
an eligible maximal simplex $\sigma_0$, the moves applied to loops in
$\Gmax{q}(\Delta)$ based at $\sigma_0$ define the group
$A^q_1(\Delta,\sigma_0)$. Applying Theorem~\ref{thm:BL} to
$G=\Gmax{q}(\Delta)$ gives its description by filled triangles and
squares.

The graph $\Gamma^q(\Delta)$ on all eligible simplices and its subgraph
on the maximal ones have the same component count and the same
$A$-fundamental groups, so nothing is lost by working with maximal
simplices. To see why, assign each eligible simplex to a maximal simplex
containing it, fixing the maximal simplices themselves. Near simplices
are sent to near or equal simplices because their common $q$-face is
retained. Every simplex is also near or equal to its assigned maximal
simplex. Replacing each simplex of a loop by its assigned maximal simplex
therefore changes the loop by a homotopy of one step. Thus the
assignment and inclusion are inverse graph homotopy equivalences. We use
this reduction throughout. The group description in terms of maximal
simplices is also given in Barcelo and Smith~\cite[Sec.~2]{BS2008}.

The fan $F_n$ shows how the connectivity graph can have a loop where
the complex has none. It is the cone on $C_n$, triangulated by joining
the cycle to one hub, with $n\ge4$: a ring of $n$ triangles around the
hub, each sharing an edge with the next. As a space it is a disc and
is contractible, but its triangles, joined when they share an edge,
form the cycle $\Gmax{1}(F_n)=C_n$ (Fig.~\ref{fig:primer}d,~e). Hence
$A^1_1(F_4)=1$ and $A^1_1(F_n)\cong\ZZ$ for $n\ge5$. These groups
describe connectivity among the maximal triangles. They are not the
ordinary fundamental groups of the fan spaces. The same effect appears
in ego networks in Sections~\ref{sec:levels} and~\ref{sec:ego}.

\section{Persistent \texorpdfstring{$A$}{A}-homotopy}
\label{sec:persistent}

\subsection{Varying the allowed cycle relations}

Fix a connected graph $G$ and a base vertex $v_0$. For an integer $k\ge2$,
let $N_{\le k}$ be the normal closure in $\pi_1(G,v_0)$ of all simple
cycles of length at most $k$, based by paths as above. Define
\[
 \Ak{k}(G,v_0)=\pi_1(G,v_0)/N_{\le k}.
\]
Increasing $k$ admits more relations, giving surjective homomorphisms
\[
 \Ak{2}(G,v_0)\twoheadrightarrow\Ak{3}(G,v_0)
 \twoheadrightarrow\Ak{4}(G,v_0)\twoheadrightarrow\cdots.
\]
This sequence, including its maps, is the group-valued construction studied
here. It is eventually trivial because the simple cycles generate the
fundamental group of a graph. It need not be abelian at intermediate scales.

The cell model is useful both for interpretation and for computation. Let
$Y_k(G)$ be $G$ with a 2-cell attached along every simple cycle of length at
most $k$. Then $Y_k(G)\subseteq Y_{k+1}(G)$ and
$\pi_1(Y_k(G),v_0)\cong\Ak{k}(G,v_0)$. To admit a relation is to declare
that a traversal of that cycle can be removed from a loop. It does not
mean adding a network edge. We use \emph{coordination scale} for $k$ in
this sense. Whether a particular relation represents a feasible physical
rearrangement depends on the application.

Abelianisation forgets the order of concatenation of loops. Reduction
modulo two also removes orientation signs and retains only parity. It gives
\begin{equation}
 V_k(G)=H_1(G;\FF)/W_k(G)
 \cong H_1(Y_k(G);\FF)
 \cong \Ak{k}(G,v_0)^{\mathrm{ab}}\otimes_{\ZZ}\FF,
 \label{eq:linearisation}
\end{equation}
where $W_k(G)$ is the span of the edge-incidence vectors of the simple
cycles of length at most $k$. Write $b_k(G)=\dim_{\FF}V_k(G)$. These
vector spaces and their quotient maps form a persistence module. For a
disconnected graph we take the direct sum of this construction over its
components, and for an empty graph it is zero.

All maps in the $k$ direction are surjective, so no new linear classes
are born after $k=2$. The module is described by intervals $[2,d)$, with
integer death scales $d\ge3$. We call the multiset of deaths
$\mathcal D(G)$ the \emph{resolvability spectrum}. It is the local
resolvability spectrum of~\cite{Rajkovic2026b}. The intervals describe
a decomposition of the vector-space module, not a canonical collection
of physical loops. Throughout the paper, a barcode associated with
persistent $A$-homotopy means this mod-two summary. No interval decomposition
of the non-abelian group sequence is asserted.

\subsection{A computable spectrum}

A minimum cycle basis is a basis of the graph cycle space over $\FF$
whose total cycle length is minimal. Algorithms for finding such bases
are established~\cite{Horton1987,Kavitha2009}. The following description is
also proved in~\cite[Section~3.2]{Rajkovic2026b}. We include the short
exchange argument to keep the paper self-contained.

\begin{samepage}
\begin{proposition}
\label{prop:mcb}
The resolvability spectrum is the multiset of lengths in any minimum
cycle basis of $G$. In particular,
\[
 b_k(G)=\#\{d\in\mathcal D(G)\mid d>k\}.
\]
\end{proposition}

\begin{proof}
Let $B$ be a minimum cycle basis. Its cycles of length at most $k$ are
independent elements of $W_k$. If they did not span $W_k$, some simple
cycle $C$ of length at most $k$ would have an expansion in $B$ involving a
cycle $D$ longer than $k$. Replacing that $D$ by $C$ would give a basis
of smaller total length. This contradicts minimality. The short basis
cycles therefore form a basis of $W_k$, and quotienting the cycle space
leaves exactly the basis elements longer than $k$. The successive rank
drops determine the death multiplicities, proving the assertion for any
minimum basis.
\end{proof}
\end{samepage}

For a graph with a cycle, define the \emph{coordination horizon}
$\Lambda(G)=\max\mathcal D(G)$, and set $\Lambda(G)=0$ for a forest.
This is the first scale at which all the mod-two classes have vanished,
with the stated convention for an initially empty spectrum. It is not
necessarily the first scale at which the full group is trivial.
For a connected graph with a cycle,
\begin{equation}
 \Lambda(G)\le 2\,\operatorname{rad}(G)+1
 \le 2\,\operatorname{diam}(G)+1.
 \label{eq:horizon}
\end{equation}
Indeed, a breadth-first spanning tree rooted at a central vertex supplies
a fundamental cycle basis with each cycle of length at most
$2\operatorname{rad}(G)+1$. Proposition~\ref{prop:mcb} then gives the
bound. For a disconnected graph apply it in each component.

The spectrum need not be insensitive to a small number of edge edits.
For example, adding an opposite chord to $C_{2m}$, with $m\ge3$, changes
its death multiset from $[2m]$ to $[m+1,m+1]$. Thus a single edge can change
long death scales substantially. Stability therefore has to be measured
against the cycles that an edit creates. The companion
paper~\cite[Section~5]{Rajkovic2026b} proves that inserting an edge whose
endpoints were at distance $\ell$ changes the barcode by at most $\ell-1$ in
bottleneck distance, and that adding a bridge leaves it unchanged. In the
example above the chord joins vertices at distance $m$, and the bound $m-1$
is attained.

\subsection{The clique and \texorpdfstring{$A$}{A}-homotopy stages}

\begin{samepage}
\begin{proposition}
\label{prop:slices}
For a connected graph $G$, with basepoint suppressed in the notation,
\begin{enumerate}
\item[(i)] $\Ak{2}(G)=\pi_1(G)$ and $b_2=\beta_1(G)$.
\item[(ii)] $\Ak{3}(G)\cong\pi_1(X(G))$ and
$b_3=\dim_{\FF}H_1(X(G);\FF)$.
\item[(iii)] $\Ak{4}(G)\cong A_1(G)$ and
$b_4=\dim_{\FF}(A_1(G)^{\mathrm{ab}}\otimes_{\ZZ}\FF)$.
\end{enumerate}
Consequently $b_2\ge b_3\ge b_4\ge\cdots$.
\end{proposition}

\begin{proof}
A simple graph has no simple cycles of length at most two. At length
three, the attached cells form the 2-skeleton of the clique complex,
and the simplices of higher dimension do not affect its fundamental group
or first homology. The length-four statement is Theorem~\ref{thm:BL}.
The dimension inequalities follow from the quotient maps.
\end{proof}
\end{samepage}

The drop at three counts independent triangle relations, not all triangles
in the graph. The drop $b_3-b_4$ counts the additional independent square
relations after triangles have been filled. We call these classes locally
resolvable at the $A$-homotopy stage. A class surviving at $k=4$ certifies a
non-trivial element of $A_1(G)$, but its dimension counts independent
mod-two classes rather than group elements or a complete set of generators.

\begin{figure}[t]
\centering
\includegraphics[width=\textwidth]{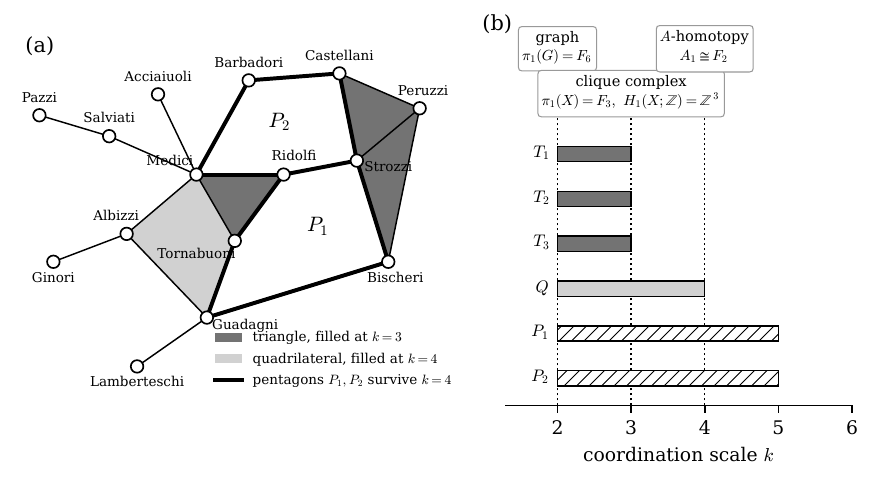}
\caption{\textbf{The Florentine marriage network.} (a) Three triangles are filled at $k=3$. The chordless square Albizzi--Medici--Tornabuoni--Guadagni supplies the further independent relation at $k=4$. The two pentagons survive this stage. (b) The mod-two coordination barcode has three deaths at three, one at four and two at five. The exact group labels follow from free-edge collapses. Integral first homology is displayed explicitly for the clique complex. The bars alone do not establish the group isomorphisms.}
\label{fig:florentine}
\alttext{Panel (a) shows the Florentine marriage network with family names at the vertices. Three triangles are shaded dark grey, the quadrilateral Albizzi--Medici--Tornabuoni--Guadagni is shaded light grey, and two pentagons, labelled $P_1$ and $P_2$, are drawn with heavy lines. Panel (b) shows a horizontal barcode against coordination scale $k$ from $2$ to $6$: three dark bars $T_1$, $T_2$ and $T_3$ end at $3$, a light bar $Q$ ends at $4$, and two hatched bars $P_1$ and $P_2$ end at $5$. Boxes above the bars give $\pi_1(G)=F_6$ for the graph, $\pi_1(X)=F_3$ and $H_1(X;\mathbb{Z})=\mathbb{Z}^3$ for the clique complex, and $A_1\cong F_2$ for $A$-homotopy.}
\end{figure}

Figure~\ref{fig:florentine} illustrates the distinction in the Florentine
marriage network~\cite{Padgett1993}. Its spectrum is
$[3,3,3,4,5,5]$. Removing the three triangle cells through free edges leaves
a graph of cycle rank three, so $\pi_1(X(G))\cong F_3$ and
$H_1(X(G);\ZZ)\cong\ZZ^3$. Removing the chordless square as well gives
$A_1(G)\cong F_2$. The two displayed pentagons can be chosen as the free
generators. Here the exact group conclusion is supported by a collapse,
not inferred from the barcode alone.

\subsection{When the barcode determines the groups}
\label{sec:freetower}

The Florentine groups were identified by combining the barcode with
collapses. In general the barcode records only the numbers $b_k(G)$.
Whether it also determines the groups $\Ak{k}(G)$ depends on whether these
groups are free. A free group is determined up to isomorphism by its rank,
and the rank of $F_r$ is the dimension $r$ of its abelianisation reduced
modulo two.

\begin{samepage}
\begin{proposition}
\label{prop:freetower}
Suppose that $\Ak{k}(G)$ is free for every $k\ge2$. Then
$\Ak{k}(G)\cong F_{b_k(G)}$, and for every finite group $H$,
\[
 |\operatorname{Hom}(\Ak{k}(G),H)|=|H|^{b_k(G)}.
\]
Moreover, the sequence of groups and surjections is isomorphic to the
sequence $F_{b_2}\to F_{b_3}\to\cdots$ in which each map sends the first
$b_{k+1}$ elements of a basis to a basis and the remaining ones to the
identity.
\end{proposition}
\end{samepage}

\begin{proof}
The first two statements follow from Eq.~\eqref{eq:linearisation} and
from the fact that a homomorphism from a free group can send a basis to
arbitrary elements. For the last statement, let $K=\Lambda(G)$, so that $\Ak{k}(G)=1$ for
$k\ge K$, and choose bases downwards, starting from the empty basis of
$\Ak{K}(G)$. Suppose that a basis $(y_1,\ldots,y_n)$ of $\Ak{k+1}(G)$ has
been chosen, with $n=b_{k+1}$, and let $x_1,\ldots,x_m$ be any basis of
$\Ak{k}(G)$. The images of the $x_i$ generate $\Ak{k+1}(G)$. By Nielsen's
reduction method, a finite generating tuple of a free group of rank $n$
can be transformed by Nielsen transformations into a basis followed by
identity elements~\cite[Ch.~I]{LS1977}. Since any two bases are related by
Nielsen transformations, the tuple can be brought to
$(y_1,\ldots,y_n,1,\ldots,1)$. The same transformations applied to
$(x_1,\ldots,x_m)$ give a new basis of $\Ak{k}(G)$ with the stated
images.
\end{proof}

In this situation homomorphism counts into finite groups, and the whole
sequence of groups and homomorphisms, are functions of the resolvability
spectrum. Section~\ref{sec:free} tests the hypothesis of
Proposition~\ref{prop:freetower} on networks and network models, and
Section~\ref{sec:periodic} constructs networks for which it fails.

\begin{figure}[!t]
\centering
\resizebox{\textwidth}{!}{\input{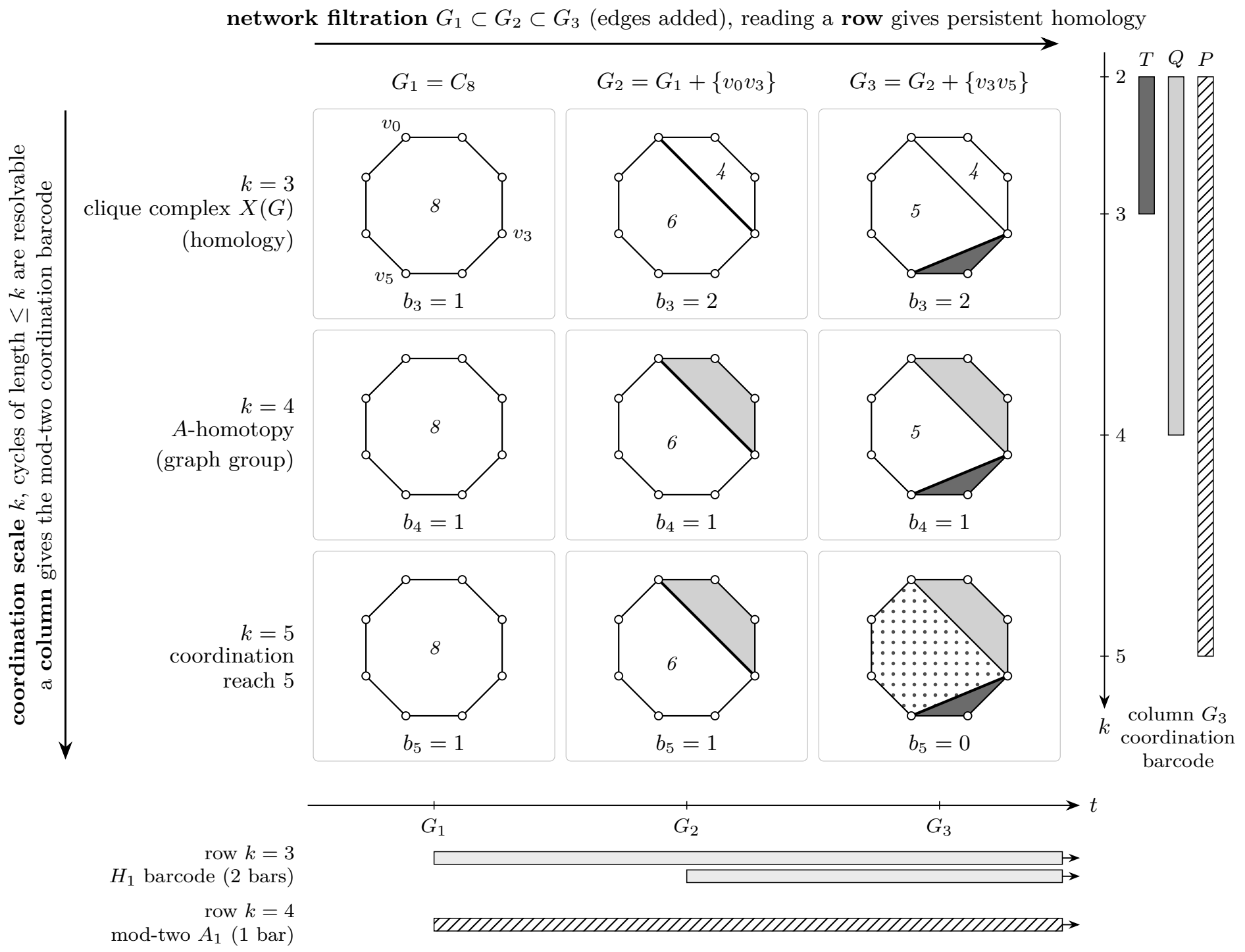}}
\caption{\textbf{Network threshold and coordination scale.} Across a row, $G_1=C_8$ acquires the chords $v_0v_3$ and $v_3v_5$. Down a column, all simple cycles of length at most $k$ are filled. Dark grey marks a triangle, light grey a square and dots a pentagon. All ranks and barcodes use $\FF$. The number $b_k$ under each panel counts the independent loop classes that survive when all simple cycles of length at most $k$ are filled, $b_k=\dim_{\FF}H_1(Y_k(G);\FF)$, as in Eq.~\eqref{eq:linearisation}. The $k=3$ row gives clique first homology persistence, with two bars surviving to $G_3$. The $k=4$ row gives one bar in the mod-two abelianised $A$-groups. Arrows extend these bars under the convention that the final graph is held fixed. For the column $G_3$, the intervals are $[2,3)$, $[2,4)$ and $[2,5)$. The drawings show enough cells to describe first homology and fundamental groups, not a claim about higher homology of the full cycle-filling complexes.}
\label{fig:twoparam}
\alttext{A grid of nine panels with three rows and three columns. The columns show the graphs $G_1$, an eight-cycle, $G_2$, which adds the chord $v_0v_3$, and $G_3$, which also adds the chord $v_3v_5$. The rows show coordination scales $k=3$, $4$ and $5$. In each panel the filled cells are shaded: the triangle in dark grey, the square in light grey and, at $k=5$ in $G_3$, the pentagon dotted. The values printed under the panels are $b_3=1,2,2$ in the first row, $b_4=1,1,1$ in the second and $b_5=1,1,0$ in the third. A vertical barcode on the right shows the three intervals of $G_3$, ending at $3$, $4$ and $5$. Horizontal bars below the grid show two bars for the $k=3$ row and one bar for the $k=4$ row.}
\end{figure}

\subsection{Two filtration parameters}

Let $G_s\subseteq G_t$ for $s\le t$ be a nested family of graphs. The
inclusion sends a cycle of length at most $k$ to one of the same length,
so it induces a map
\[
 V_{s,k}=H_1(G_s;\FF)/W_k(G_s)
 \longrightarrow V_{t,k}=H_1(G_t;\FF)/W_k(G_t).
\]
These maps commute with the quotient maps obtained by increasing $k$.
The spaces $V_{t,k}$ and these maps form a two-parameter persistence module.
Its dimensions $b_k(G_t)$ are a summary, not the module itself. Equivalently,
it is the first homology module of the bifiltration $Y_k(G_t)$. The group
version uses compatible basepoints in the relevant components, whereas homology
does not need such choices. With the overlap threshold of a simplicial complex
in place of the network filtration, the analogous module is constructed
in~\cite[Section~6]{Rajkovic2026b}.

Figure~\ref{fig:twoparam} separates the two parameters. Across the page,
edges are added to a network. Down the page, the network is held fixed
and more loop relations are imposed. A class can persist across network
thresholds and still disappear as soon as square moves are admitted.

For fixed $k=3$, the horizontal module is naturally the first homology
persistence of the clique complexes $X(G_t)$. For $k=4$, it is the
persistence of the mod-two abelianised $A$-groups. For fixed $t$, the
vertical module measures response to the length of permitted relations.
In Fig.~\ref{fig:twoparam}, the $k=3$ row has two bars that survive to the last
graph, whereas the $k=4$ row has one. The extra class in $G_2$ is removed
by a square relation. The example shows why persistence across network
thresholds does not determine persistence in coordination scale.

Weighted networks supply nested graphs by thresholding, with the
inequality direction stated according to the meaning of the weight.
A time sequence with both edge additions and deletions is generally not
nested and needs a different construction or an explicitly cumulative
network. No general two-dimensional barcode is assumed here.

\subsection{Connectivity of cliques inside an ego network}
\label{sec:levels}

There is a separate choice of graph before any cycle relations are
imposed. We can work on the original network $G$, or on
$\Gmax{q}(X(G))$, whose vertices represent overlapping maximal cliques.
The resulting invariants concern different objects. In particular,
the inequality $b_4\le b_3$ does not compare the original graph with
its clique-connectivity graph.

\begin{samepage}
\begin{proposition}
\label{prop:ego}
Let $E_v$ be the induced subgraph on a vertex $v$ and its neighbours.
Then $X(E_v)$ is the closed star of $v$ in $X(G)$ and is a cone.
Thus $\widetilde H_j(X(E_v);\FF)=0$ for every $j$ and
$\pi_1(X(E_v))=1$. If $v$ has a neighbour, the vertices of
$\Gmax{1}(X(E_v))$ are the maximal cliques of $G$ containing $v$,
and two of them are adjacent when they share a vertex other than $v$.
\end{proposition}

\begin{proof}
Every clique in $E_v$ extends to one containing $v$, so its clique
complex is a cone with apex $v$. Its maximal cliques are exactly the
maximal cliques of $G$ through $v$. Two of them share an edge precisely
when their intersection contains another vertex besides $v$.
\end{proof}
\end{samepage}

For an isolated $v$, there is no eligible maximal simplex at $q=1$.
For other egos the connectivity graph can be disconnected. Its
non-trivial $A$-classes describe loops of overlapping cliques, even though
the ego clique complex itself is contractible. The fan examples show this
without any data. Section~\ref{sec:ego} gives network instances. Their
cochain certificates are homological tests on $Y_4(\Gmax{1}(X(E_v)))$,
so they do not contradict the vanishing homology of $X(E_v)$.

\subsection{Computation}

The numerical spectra use minimum cycle bases, with the
\texttt{igraph} implementation~\cite{igraph}. Direct elimination of
triangle and square incidence vectors over $\FF$ independently checks
$b_3$ and $b_4$. Selected spectra are also checked with
\texttt{NetworkX}~\cite{networkx}. We retain the distinction between a
linear rank calculation and a certificate about the full group.

The accompanying code provides the graph preprocessing, spectra,
short-cycle ranks, ego scans and certificate files. A MATLAB reference
implementation computes a minimum cycle basis by the support-vector
method described in~\cite{Kavitha2009}, using shortest paths in a two-sheeted
parity graph. It is intended for small and moderate examples rather than
as a replacement for the compiled routine on the largest networks.
The package also includes MATLAB routines for short-cycle ranks,
maximal-clique connectivity and inspection of the saved spectra.

\section{Free groups in networks}
\label{sec:free}

Proposition~\ref{prop:freetower} reduces the question of
Section~\ref{sec:intro} to freeness. This section gives a test for it,
applies the test to the benchmark networks and to network models, and
explains why the groups of these networks are free.

\subsection{A freeness test}
\label{sec:freetest}

Every simple cycle of $G$ lies in a single biconnected block. The complex
$Y_k(G)$ is therefore obtained from the complexes $Y_k(B)$ of the blocks
$B$ by joining them at cut vertices in a tree pattern, and the
Seifert--van Kampen theorem gives a free product
\[
 \Ak{k}(G)\cong\mathop{\ast}_B \Ak{k}(B)
\]
over the blocks that contain a cycle~\cite[Thm.~1.20 and
Example~1.21]{Hatcher2002}. It suffices to treat each block. Two
reductions keep the presentations small.

A simple cycle $C$ is \emph{isometric} if the distance between any two
of its vertices along $C$ equals their distance in $G$.

\begin{lemma}
\label{lem:isometric}
For every $k\ge3$, $N_{\le k}$ is the normal closure of the isometric
cycles of length at most $k$.
\end{lemma}

\begin{proof}
A closed walk of length $m$ that is not a simple cycle can be cut at a
repeated vertex, or shortened by a backtrack, into closed walks of smaller
length. Its class is therefore a product of conjugates of simple cycles of
length at most $m$. Now let $C$ be a simple cycle of length $m\le k$ that
is not isometric. Choose vertices $u$ and $v$ of $C$ with
$d_G(u,v)<d_C(u,v)$, and a shortest path $P$ from $u$ to $v$ in $G$. The
vertices $u$ and $v$ divide $C$ into an arc $\alpha$ from $u$ to $v$ of
length $a=d_C(u,v)$ and an arc $\alpha'$ from $v$ back to $u$ of length
$m-a\ge a$. Up to backtracks, $C$ is the product of the closed walks
$\alpha\bar P$ and $P\alpha'$, where $\bar P$ is $P$ reversed. Their
lengths are $a+d_G(u,v)<2a\le m$ and $d_G(u,v)+m-a<m$. Induction on $m$
shows that $C$ lies in the normal closure of isometric cycles of length at
most $k$. The converse inclusion holds because isometric cycles are simple
cycles.
\end{proof}

\begin{lemma}
\label{lem:dominated}
Let $v$ be a vertex with a neighbour $w$ such that every neighbour of $v$
other than $w$ is also a neighbour of $w$. Then
$\Ak{k}(G)\cong\Ak{k}(G-v)$ for every $k\ge3$, with any base vertex
other than $v$.
\end{lemma}

\begin{proof}
Let $r$ send $v$ to $w$ and fix all other vertices. It maps each edge of
$G$ to an edge of $G-v$, except the edge $vw$, which goes to the vertex
$w$. A closed walk of length $m$ is therefore sent to a closed walk of
length at most $m$. By the first step of the proof of
Lemma~\ref{lem:isometric}, $r$ induces a homomorphism
$\rho\colon\Ak{k}(G)\to\Ak{k}(G-v)$. The inclusion of $G-v$ induces
$\iota\colon\Ak{k}(G-v)\to\Ak{k}(G)$, and $\rho\circ\iota$ is the
identity. Replacing one passage $x,v,y$ of a loop by $x,w,y$ changes it by
the closed walk $(x,v,y,w,x)$. This walk is a product of the triangles
$xvw$ and $vyw$, or it reduces to one of them or to a backtrack when $x$
or $y$ equals $w$. Hence $\iota\circ\rho$ is the identity for $k\ge3$.
\end{proof}

In a clustered network many vertices satisfy the hypothesis of
Lemma~\ref{lem:dominated}, often with a hub as $w$, and repeated deletion
can shrink a graph considerably. The groups for $k\ge3$, and hence $b_k$
for $k\ge3$, are unchanged.

For a block $B$ we take as generators the edges outside a breadth-first
spanning tree, and one relator for each isometric cycle of length at most
$k$, written in these generators. We then remove generators. If a
generator occurs exactly once in a relator, that relator can be solved for
it. Substituting the solution into the other relators and deleting the
generator and the relator gives a presentation of the same group. Relators
that reduce to the empty word are deleted, and so are repetitions up to
cyclic permutation and inversion. When every relator has been removed, the
group is free on the remaining generators, and the sequence of
eliminations is a proof. This extends the free-edge collapse of
Lemma~\ref{lem:cert}(c), which removes the generator of a free edge.

When relators remain, three tests can prove that the group is not free.
Let $r$ be the rank of its integral first homology and $g$ the number of
remaining generators.
\begin{enumerate}
\item[(i)] Torsion in the integral first homology, found from the Smith
normal form of the exponent-sum matrix, excludes a free group.
\item[(ii)] A free group with first homology $\ZZ^r$ is $F_r$, and it has
$6^r$ homomorphisms to the symmetric group $S_3$. We count these
homomorphisms for the remaining presentation by exhaustive search. A
different count excludes a free group.
\item[(iii)] Suppose that every remaining relator has exponent sum zero in
each generator, so that $r=g$. A free group would then be $F_g$, and the
presentation would give a surjection $F_g\to F_g$ whose kernel contains
the remaining relators, which are non-empty reduced words.
This is impossible because finitely generated free groups are
Hopfian~\cite[Ch.~I]{LS1977}.
\end{enumerate}
A case decided by none of these would be reported as undecided. None
occurred among the graphs reported below. As a check, at every $k$ the
mod-two dimension of the first homology of the final presentation was
compared with $b_k(G)$ computed independently from the ranks of the
cycle spaces. For the benchmarks it was also compared with
the spectra in Table~\ref{tab:atlas}. All values agreed. For dense graphs the cycles of
length $k+1$ are rewritten in the reduced presentation obtained at $k$
instead of starting again from the spanning tree. The test recovers the torus group of
the periodic square grid in Proposition~\ref{prop:lattices}(d), where
test~(iii) applies. On grids of
$6\times6$ squares with the identifications of a Klein bottle and of a
projective plane, test~(i) detects the torsion of $\ZZ\oplus\ZZ/2$ and of $\ZZ/2$.

\subsection{Networks and models tested}
\label{sec:freecomplex}

Table~\ref{tab:data} lists the nine benchmark networks, which are analysed
further in Section~\ref{sec:complex}. We discard edge weights and
directions, remove self-loops and merge parallel edges. We then retain
the largest connected component. This treats the data as structural
examples. In a directed or weighted physical model, the admissible loop
relations would require a separate definition. The data were obtained
from NetworkX~\cite{NXdata}, from Newman's network data
collection~\cite{NewmanData} and from the Gephi sample
datasets~\cite{GephiData}. Table~\ref{tab:data} gives the source of each
network. Exact retrieval locations, checksums, labels and the resulting
edge lists accompany the code.

The diseasome file is the distributed mixed network containing both
disease--gene associations and disease--disease links. It is not just the
bipartite relation of the original study~\cite{Goh2007}. The airline file
is a Gephi sample with an unidentified original source~\cite{GephiData}.
We retain it as a computational illustration and do not use it to infer
properties of air transport in general.

\begin{table}[!htbp]
\centering
\small
\caption{\textbf{Benchmark networks.} $V$ and $E$ refer to the largest
connected component.}
\label{tab:data}
\medskip
\begin{tabular}{@{}>{\raggedright}p{2.5cm}>{\raggedright}p{5.2cm}rrl@{}}
\toprule
Network & Nodes and edges & $V$ & $E$ & Ref.\\
\midrule
Karate club & members of a university karate club; interactions outside the club & 34 & 78 & \cite{Zachary1977,NXdata}\\
Les Misérables & characters of Hugo's novel; appearance in a common chapter & 77 & 254 & \cite{Knuth1993,NXdata}\\
Florentine families & Renaissance Florentine families; marriage ties (the isolated Pucci family omitted) & 15 & 20 & \cite{Padgett1993,NXdata}\\
College football & U.S.\ Division I-A college teams; games played in the 2000 season & 115 & 613 & \cite{GN2002,NewmanData}\\
Network science & scientists working on networks; coauthorship & 379 & 914 & \cite{Newman2006,GephiData}\\
\emph{C.\ elegans} & neurons of the nematode; synaptic connections & 297 & 2148 & \cite{White1986,WS1998,GephiData}\\
Diseasome & 516 human diseases and 903 disease genes; 1550 disease--gene associations and 1188 disease--disease links & 1419 & 2738 & \cite{Goh2007,GephiData}\\
Airlines & airports; air routes (sample network distributed with \textsc{Gephi}) & 235 & 1297 & \cite{GephiData}\\
US power grid & generators, transformers and substations of the western U.S.; high-voltage lines & 4941 & 6594 & \cite{WS1998,GephiData}\\
\bottomrule
\end{tabular}
\end{table}

Table~\ref{tab:free} lists all graphs tested, the benchmarks among them. For every graph and every
$k$ up to the scale at which the group becomes trivial, the eliminations
removed all relators. The groups are therefore free, and
Proposition~\ref{prop:freetower} applies to each graph.

\begin{table}[!htbp]
\centering
\small
\caption{\textbf{Graphs for which every group $\Ak{k}(G)$ was certified free.} The last column gives the range of $k$ examined, from $3$ to the scale at which the group becomes trivial; for the $G(n,p)$ graphs only $k=3$ was examined. Randomisations use double-edge swaps with ten times as many swaps as edges and retain the largest component. Graphs were generated with NetworkX~3.6.1 and the scripts provided, with fixed seeds. The configuration-model degrees have minimum two and are capped at $2\sqrt n$; loops and multiple edges are removed.}
\label{tab:free}
\medskip
\footnotesize
\setlength{\tabcolsep}{4pt}
\begin{tabular}{@{}>{\raggedright}p{3.3cm}>{\raggedright}p{5.3cm}rr@{}}
\toprule
Graphs & Parameters & Number & $k$\\
\midrule
Benchmark networks & Table~\ref{tab:data} & 9 & 3--31\\
Maximal-clique connectivity graphs & $\Gmax{q}(X(G))$ of the benchmarks, $q=0,1,2$, when they contain a cycle; $q=0$ omitted for \emph{C.\ elegans} & 22 & 3--31\\
Randomised benchmarks & three per benchmark & 27 & 3--17\\
Watts--Strogatz~\cite{WS1998} & $n=1000$, $k_0=10$, $p=1, 0.5, 0.1$ (two each), $p=0.01$ (one) & 7 & 3--21\\
Barab\'asi--Albert~\cite{BA1999} & $n=2000$, $m=2,3,5$ & 3 & 3--9\\
Holme--Kim~\cite{HolmeKim2002} & $n=2000$, $(m,p_t)=(3,0.5),(5,0.9)$ & 2 & 3--7\\
Configuration model~\cite{MolloyReed1995} & $n=3000$, degrees with power-law tail, exponents $2.5$ and $2.2$ & 2 & 3--9\\
Hyperbolic random graphs~\cite{Krioukov2010} & temperature zero; $n=1000$, exponent $2.5$, mean degree $10$; $n=2000$, exponent $2.2$, mean degree $6$ & 2 & 3--5\\
LFR benchmark~\cite{LFR2008} & $n=1000$, mixing $\mu=0.2$ & 1 & 3--6\\
Stochastic block model~\cite{Holland1983} & four blocks of $250$, $p_{\mathrm{in}}=0.05$, $p_{\mathrm{out}}=0.002$ & 1 & 3--6\\
Duplication--divergence~\cite{Ispolatov2005} & $n=2000$, retention probability $0.4$ & 1 & 3--4\\
Random geometric graph~\cite{Penrose2003} & unit square, $n=800$, radius $0.07$ & 1 & 3--12\\
$G(n,p)$~\cite{ErdosRenyi1960} & $(n,p)=(500,0.10)$, $(700,0.11)$, $(1000,0.075)$, $(1000,0.085)$, $(1000,0.095)$, $(1500,0.075)$ & 6 & 3\\
Street networks~\cite{OSMHelsinki,OSMOakland} & Helsinki driving and walking networks, a second Helsinki extract, West Oakland, and three randomisations of each & 16 & 3--27\\
\bottomrule
\end{tabular}
\end{table}

For the benchmarks this identifies groups that Table~\ref{tab:atlas} of
Section~\ref{sec:benchmarks} records only through mod-two ranks. The $A$-group is $\ZZ$ for
\emph{C.\ elegans}, $F_7$ for football, $F_2$ for the Florentine
families, in agreement with Section~\ref{sec:persistent}, $F_6$ for the
coauthorship network, $F_{10}$ for the diseasome and $F_{811}$ for the
power grid. For the maximal-clique connectivity graphs
$\Gmax{1}(X(G))$, every component has a
free $A$-group, and the ranks add up to three for football and to one for
each of the diseasome and the power grid. The non-trivial components of
the Tennessee and colon-cancer egos of Section~\ref{sec:ego} have
$A$-group $\ZZ$, as for Bowling Green.

Two street networks were included because their graphs are close to
planar. The street data are OpenStreetMap extracts distributed with the
\texttt{pyrosm} package (Helsinki) and with the test data of the
\texttt{OSMnx} package (West Oakland). Intersections and dead ends are the
vertices. Chains of degree-two vertices are contracted to single edges,
and loops and parallel edges are discarded. The Helsinki driving network
has 161 vertices and 224 edges and is planar. The Helsinki walking
network has 2440 vertices and 3474 edges and is not planar.

\subsection{Why these groups are free}
\label{sec:whyfree}

Two mechanisms account for much of Table~\ref{tab:free}. The first
applies to sparse networks, whose short cycles rarely overlap.

\begin{lemma}
\label{lem:disjoint}
Let $G$ be connected and suppose that its simple cycles of length at most
$k$ are pairwise vertex-disjoint. Then $\Ak{k}(G)$ is free of rank
$\beta_1(G)-s$, where $s$ is the number of these cycles.
\end{lemma}

\begin{proof}
Delete one edge from each of the $s$ cycles. The remaining edges of the
cycles form disjoint paths, so they lie in a spanning tree $T$. With
respect to $T$, each short cycle is written as the single generator of
its deleted edge, and distinct cycles give distinct generators. The
quotient of a free group by a subset of a basis is free on the remaining
basis elements.
\end{proof}

\begin{proposition}
\label{prop:sparse}
Fix $k\ge3$ and $c>0$, and let $G$ be an Erd\H{o}s--R\'enyi graph
$G(n,c/n)$. With probability tending to one as $n\to\infty$, any two
distinct simple cycles of $G$ of length at most $k$ are vertex-disjoint,
and hence $\Ak{k}(H)$ is free for every component $H$ of $G$.
\end{proposition}

\begin{proof}
If two distinct simple cycles of length at most $k$ share a vertex, their
union is connected, has $v\le2k-1$ vertices and has cycle rank at least
two. It therefore contains a connected subgraph with the same $v$
vertices and exactly $v+1$ edges, a spanning tree with two further edges.
The expected number of such subgraphs in $G(n,c/n)$ is at most
\[
 \sum_{v=4}^{2k-1} n^v\binom{\binom v2}{v+1}\Bigl(\frac cn\Bigr)^{v+1}
 =O\Bigl(\frac1n\Bigr),
\]
and the claim follows from Markov's inequality and
Lemma~\ref{lem:disjoint} applied to each component.
\end{proof}

Proposition~\ref{prop:sparse} concerns a fixed $k$ as the graph grows.
Table~\ref{tab:free} covers all $k$ for the graphs tested, including
lengths comparable with the diameter, where it gives no guarantee. The
second mechanism applies to clustered networks. There the short cycles
lie in dense neighbourhoods, often around a hub that is adjacent to every
vertex of the cycle and fills it with triangles. Lemma~\ref{lem:dominated}
makes this quantitative. Repeated deletion of dominated vertices reduces
the hyperbolic graph with mean degree $10$ from $953$ to $15$ vertices,
and the one with mean degree $6$ from $1303$ vertices to a single vertex,
so all its groups are trivial. In the Holme--Kim graph with $p_t=0.9$ it
removes $778$ of $2000$ vertices.

Freeness is not a general property of random graphs. Babson shows that
the clique complex of $G(n,p)$ is not simply connected for edge
probabilities between $n^{-1/2}$ and $n^{-1/3}$, within the precise
bounds stated in~\cite{Babson2012}, and notes that $n^{-1/2}$ is a
threshold for the vanishing of first homology with $\FF$ coefficients.
Combining the two statements, in this range $\Ak{3}(G)$ is non-trivial
while $b_3(G)=0$ with probability tending to one. A non-trivial free
group has non-zero mod-two first homology, so $\Ak{3}(G)$ is then not free
and the barcode records nothing about it. Costa, Farber and Horak study
further properties of these groups~\cite{CFH2015}. Our $G(n,p)$ graphs in
this range, with at most $1500$ vertices, still had $b_3$ between $8$ and
$135$ and free groups, so they had not reached the asymptotic regime.
We also do not know whether $\Ak{k}(G)$ is free for every planar graph.
The planar Helsinki driving network supports this, as do $1500$ random
planar graphs obtained by thinning and subdividing Delaunay
triangulations, but we have no proof.

\section{Networks with periodic structure}
\label{sec:periodic}

Section~\ref{sec:free} found free groups for every graph in
Table~\ref{tab:free}. The
constructions in this section are different. In each of them short cycles
close up into a torus through two independent periodic directions, and the
groups are not free. The periodic square lattice is the basic case. A
random spatial network on a torus, a lattice with random shortcuts and a
network repeated over periodic time contain the same structure.

\subsection{Lattices}
\label{sec:lattices}

\begin{figure}[t]
\centering
\includegraphics[width=\textwidth]{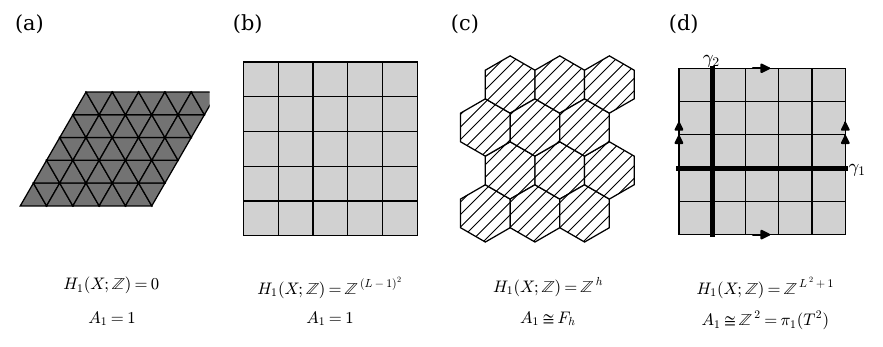}
\caption{\textbf{Lattice examples.} Labels use integral homology of the clique complex and the graph $A$-fundamental group. The triangular and honeycomb patches are discs before their displayed faces are removed. Square shading indicates cells filled for $A$-homotopy, not faces of the clique complex. In (d), opposite sides are identified, and the displayed fundamental domain has five plaquettes along each direction, so the periodic graph satisfies $L\ge5$. The loops $\gamma_1,\gamma_2$ generate the torus fundamental group. The honeycomb hatching marks unfilled faces.}
\label{fig:lattices}
\alttext{Four lattice patches with labels beneath them. Panel (a) is a triangular lattice with all triangles shaded dark grey, labelled $H_1(X;\mathbb{Z})=0$ and $A_1=1$. Panel (b) is a square grid with light grey squares, labelled $H_1(X;\mathbb{Z})=\mathbb{Z}^{(L-1)^2}$ and $A_1=1$. Panel (c) is a patch of hatched hexagons, labelled $H_1(X;\mathbb{Z})=\mathbb{Z}^h$ and $A_1\cong F_h$. Panel (d) is a square grid with arrows marking identified opposite sides and two heavy loops, $\gamma_1$ horizontal and $\gamma_2$ vertical, labelled $H_1(X;\mathbb{Z})=\mathbb{Z}^{L^2+1}$ and $A_1\cong\mathbb{Z}^2=\pi_1(T^2)$.}
\end{figure}

Regular lattices give examples for which both the groups and their linear
summaries can be computed directly. They also show why the rule for filling
cells matters (Fig.~\ref{fig:lattices}).

\begin{samepage}
\begin{proposition}
\label{prop:lattices}
Use integral coefficients in the homology groups below.
\begin{enumerate}
\item[(a)] For a finite triangular lattice patch whose filled triangular
faces form a disc, $H_1(X;\ZZ)=0$ and $A_1=1$.
\item[(b)] For the open $L\times L$ square grid, with $L\ge2$ vertices
along each side, $H_1(X;\ZZ)\cong\ZZ^{(L-1)^2}$ and $A_1=1$.
\item[(c)] For the graph of a honeycomb disc patch with exactly $h$
bounded faces, all hexagons, $H_1(X;\ZZ)\cong\ZZ^h$ and $A_1\cong F_h$.
\item[(d)] For the periodic grid $C_L\mathbin{\square}C_L$, with $L\ge5$,
$H_1(X;\ZZ)\cong\ZZ^{L^2+1}$ and $A_1\cong\ZZ^2=\pi_1(T^2)$.
\end{enumerate}
\end{proposition}

\begin{proof}
In (a), the clique complex is the triangular disc, so triangle relations
already kill the fundamental group. Adding square relations cannot change
a trivial group. In (b), there are no triangles and the cycle rank is
$(L-1)^2$. The squares are precisely the plaquettes, whose attachment gives
a disc. In (c), there are no triangles or squares. Euler's formula gives
cycle rank $h$, and the $A$-group remains the free fundamental group of
the graph. In (d), the graph has $L^2$ vertices and $2L^2$ edges, hence
cycle rank $L^2+1$. For $L\ge5$ its four-cycles are exactly the plaquettes.
Filling them gives the square cell decomposition of the torus.
\end{proof}
\end{samepage}

For $L=8$, the periodic grid has 65 independent classes in clique first
homology and two after square relations are imposed. Both calculations
are correct. The clique complex keeps each square unfilled, whereas the
cell model used by $A$-homotopy includes the plaquettes. If the network is
intended to discretise a square-tiled surface, the latter choice preserves
its fundamental group. This is a statement about the chosen cell model,
not a general guarantee of recovering an underlying space from its graph.
The restriction $L\ge5$ excludes additional short wrapping cycles present
on smaller periodic grids.

\subsection{A spatial network on a torus}
\label{sec:torus}

Place $n$ points independently and uniformly in the unit square with
opposite sides identified, and join two points when their distance in
the torus metric is less than $r$. This is a random geometric
graph~\cite{Penrose2003} with periodic boundary conditions, a common
idealisation of spatial networks~\cite{Barthelemy2011}. We generated six
graphs with $n=800$ and $r=0.07$ and three with $n=1500$ and $r=0.05$, so
that the mean degree is about twelve.

In all nine graphs the groups are free at small and large $k$, but over a
window of three to eleven consecutive lengths the remaining presentation
is $\langle a,b\mid aba^{-1}b^{-1}\rangle$, so $\Ak{k}(G)\cong\ZZ^2$
(Table~\ref{tab:torus} and Fig.~\ref{fig:one}a). Throughout the window $b_k=2$, which is also the
value for a graph with two independent long cycles and group $F_2$. In six
of the nine graphs the change from $F_2$ to $\ZZ^2$ occurs at a length at
which no bar of the barcode ends. A commutator relation has zero mod-two
homology class, so its addition leaves $b_k$ unchanged. One reading of
this is that just below the window the filled cells cover the torus
except for one region whose boundary is the commutator of the two
loops around the torus, and that this region is filled at the first
length of the window. In the other three graphs the group changes from $F_3$ to $\ZZ^2$ at a
length where one bar ends. The barcode then records a death but not the
new commutation.

\begin{table}[!htbp]
\centering
\small
\caption{\textbf{Random geometric graphs on the torus.} The window lists the lengths $k$ at which $\Ak{k}(G)\cong\ZZ^2$. The column $b_{k-1},\,b_k$ gives the dimensions just below and at the first length of the window. The last column records whether the group changed from $F_2$ to $\ZZ^2$ there without a death in the barcode.}
\label{tab:torus}
\medskip
\begin{tabular}{@{}rrrcccc@{}}
\toprule
$n$ & $r$ & Seed & Window & $b_{k-1},\,b_k$ & Group below window & No death\\
\midrule
800 & 0.07 & 1 & 14--16 & 2, 2 & $F_2$ & yes\\
800 & 0.07 & 2 & 12--16 & 2, 2 & $F_2$ & yes\\
800 & 0.07 & 3 & 10--16 & 3, 2 & $F_3$ & no\\
800 & 0.07 & 4 & 11--16 & 3, 2 & $F_3$ & no\\
800 & 0.07 & 5 & 12--16 & 3, 2 & $F_3$ & no\\
800 & 0.07 & 6 & 12--16 & 2, 2 & $F_2$ & yes\\
1500 & 0.05 & 1 & 15--23 & 2, 2 & $F_2$ & yes\\
1500 & 0.05 & 2 & 13--23 & 2, 2 & $F_2$ & yes\\
1500 & 0.05 & 3 & 18--24 & 2, 2 & $F_2$ & yes\\
\bottomrule
\end{tabular}
\end{table}

\subsection{A periodic grid with shortcuts}
\label{sec:shortcuts}

Proposition~\ref{prop:lattices}(d) gives $A_1\cong\ZZ^2$ for the periodic
square grid. Random shortcuts, as in small-world models, add free
generators but do not remove the torus.

\begin{proposition}
\label{prop:shortcuts}
Let $L\ge5$ and let $G$ be obtained from $C_L\mathbin{\square}C_L$ by
adding $s$ new edges, none of which lies on a simple cycle of length at
most four in $G$. Then $A_1(G)\cong\ZZ^2\ast F_s$.
\end{proposition}

\begin{proof}
By hypothesis every simple cycle of length at most four in $G$ is a cycle
of $C_L\mathbin{\square}C_L$. By Proposition~\ref{prop:lattices}(d) these
are the plaquettes, so $Y_4(G)$ is the torus $Y_4(C_L\mathbin{\square}C_L)$
with $s$ edges attached at their end vertices. Collapsing a spanning tree
of the grid, a contractible subcomplex, is a homotopy
equivalence~\cite[Prop.~0.17]{Hatcher2002}. It turns each added edge into
a circle attached at one point, so $Y_4(G)$ is homotopy equivalent to the
wedge of a torus and $s$ circles. The fundamental group of this wedge is
$\ZZ^2\ast F_s$~\cite[Example~1.21]{Hatcher2002}.
\end{proof}

For $L=20$ we added $30$ edges between vertex pairs drawn uniformly at
random, with four seeds. In each sample all $30$ edges were new. For the
first seed no added edge lies on a triangle or square, and
Proposition~\ref{prop:shortcuts} gives $A_1(G)\cong\ZZ^2\ast F_{30}$, with
$b_4=32$. In the other three samples one, four and three added edges lie
on squares, and the proposition does not apply. In the four samples the
freeness test shows that the groups are not free from $k=4$ up to $k=9$,
$8$, $10$ and $9$, respectively. Apart from the first sample at $k=4$, we
have not identified these groups.

\subsection{A network repeated over periodic time}
\label{sec:temporal}

Temporal networks are often represented by a sequence of
snapshots~\cite{HolmeSaramaki2012}. Suppose that the same network $G$ is
observed at $T$ time steps in a periodic schedule, and join each vertex
to its own copy at the next step, with the last step joined to the first.
The result is the Cartesian product $G\mathbin{\square}C_T$. Each edge
$uv$ and its copy one step later bound a square.

\begin{proposition}
\label{prop:product}
Let $G$ and $H$ be connected graphs and $k\ge4$. Then
\[
 \Ak{k}(G\mathbin{\square}H)\cong\Ak{k}(G)\times\Ak{k}(H).
\]
\end{proposition}

\begin{proof}
For edges $gg'$ of $G$ and $hh'$ of $H$, the vertices $(g,h)$, $(g',h)$,
$(g',h')$ and $(g,h')$ form a square, which we call mixed. Attaching cells
to all mixed squares of $G\mathbin{\square}H$ gives the product cell
complex $G\times H$, whose fundamental group is $\pi_1(G)\times\pi_1(H)$,
the isomorphism being induced by the two projections~\cite[Prop.~1.12]{Hatcher2002}.
Since $k\ge4$, the mixed squares are among the cells of
$Y_k(G\mathbin{\square}H)$. That complex is therefore obtained from
$G\times H$ by attaching cells along the remaining simple cycles of length
at most $k$, and $\Ak{k}(G\mathbin{\square}H)$ is the quotient of
$\pi_1(G)\times\pi_1(H)$ by the normal closure $M$ of their
classes~\cite[Prop.~1.26]{Hatcher2002}. A cycle of length $m\le k$
projects to closed walks in $G$ and in $H$, after deletion of immediate
repetitions, whose lengths add up to $m$. By the first step of the proof of
Lemma~\ref{lem:isometric}, these walks represent elements of
$N_{\le k}(G)$ and $N_{\le k}(H)$, so $M\subseteq N_{\le k}(G)\times
N_{\le k}(H)$. Conversely, the cycles of length at most $k$ in a copy of
$G$ or of $H$ have classes that normally generate $N_{\le k}(G)\times1$
and $1\times N_{\le k}(H)$. Hence $M=N_{\le k}(G)\times N_{\le k}(H)$, and
the quotient is $\Ak{k}(G)\times\Ak{k}(H)$.
\end{proof}

The statement fails for $k=3$ in general, because the mixed squares are
then not filled. For example, $C_5\mathbin{\square}C_5$ has no triangles,
so its group at $k=3$ is free of rank $26$.
For a cycle, $\Ak{k}(C_T)\cong\ZZ$ when $k<T$ and $\Ak{k}(C_T)=1$ when
$k\ge T$. Hence, for $k\ge4$,
\[
 \Ak{k}(G\mathbin{\square}C_T)\cong
 \begin{cases}
 \Ak{k}(G)\times\ZZ, & k<T,\\
 \Ak{k}(G), & k\ge T,
 \end{cases}
 \qquad b_k(G\mathbin{\square}C_T)=b_k(G)+[k<T],
\]
where $[k<T]$ is one if $k<T$ and zero otherwise. The wedge
$G\vee C_T$, obtained by attaching a separate cycle of length $T$ at one
vertex of $G$, has the same $b_k$ for every $k\ge4$, but its group is the
free product $\Ak{k}(G)\ast\ZZ$ for $k<T$. When $\Ak{k}(G)\cong F_r$ with
$r\ge1$ and $4\le k<T$, the product $F_r\times\ZZ$ contains $\ZZ^2$ and is
therefore not free, since subgroups of free groups are
free~\cite[Ch.~I]{LS1977}, whereas $F_r\ast\ZZ\cong F_{r+1}$. The barcode
cannot distinguish a network that persists through the period from one
with an additional independent cycle. The group can, because the squares
formed by each edge and its copy make every loop of $G$ commute with the
loop that follows a vertex once around the period.

For the Florentine families and $T=8$, $\Ak{4}$ is $F_2\times\ZZ$ for the
product and $F_3$ for the wedge, with $b_4=3$ in both cases
(Fig.~\ref{fig:one}b). For $5\le
k\le7$ both groups are $\ZZ$, because the two pentagons of the Florentine
network are filled at $k=5$. For college football and $T=8$, the product
gives $F_7\times\ZZ$ at $k=4$ and $\ZZ^2$ at $k=5$. For the karate club,
whose $A$-group is trivial, the product gives $\ZZ$ for $4\le k\le7$. The
freeness test agrees with these statements. It certifies the products as
non-free exactly where the proposition gives $F_r\times\ZZ$ with $r\ge1$,
and as free elsewhere. A product is the simplest
periodic model. When the network changes during the period, the graph is
no longer a product and its groups need not split in this way. We have not
studied that case.

\section{Barcodes of network models and data}
\label{sec:complex}

For the benchmark networks, their clique-connectivity graphs and the
Watts--Strogatz graphs tested in Section~\ref{sec:free}, every group in the
sequence is free. By Proposition~\ref{prop:freetower} the barcode then
determines each group and each map, so for these networks the mod-two
spectrum describes the whole construction. The seven Watts--Strogatz graphs of
Table~\ref{tab:free} were generated separately, with the parameters of the
experiment below at four of its thirteen probabilities. For the 260 saved
realisations of that experiment, freeness is an extrapolation from them. This section reads
the spectra.

\subsection{The Watts--Strogatz model}
\label{sec:ws}

We use the rewiring construction of Watts and Strogatz~\cite{WS1998}, with
$n=1000$ vertices initially joined to their five nearest neighbours in
each direction. Thus the initial degree is $k_0=10$. The saved experiment
contains twenty realisations at each of thirteen probabilities from
$10^{-4}$ to $1$. The NetworkX generator repeats the construction when
necessary until a connected graph is obtained. We therefore study the
connected ensemble. The accompanying records contain the probability,
seed, graph parameters, clustering coefficient, mean shortest-path length
and death multiset for each realisation.

At $p=0$, the graph is $C_{1000}^5$. Its clique complex has dimension five,
because six consecutive vertices form a clique, but is homotopy equivalent
to a circle~\cite[Prop.~2.2]{Adamaszek2013}. Its spectrum has 4000 deaths
at three and one at 200. One can verify this without a large computation.
Triangle moves expand every edge into consecutive unit edges. No triangle
winds around the ring, so the quotient by triangle relations is infinite
cyclic. A cycle of nonzero winding needs at least 200 edges, since an edge
advances at most five positions, and steps of five give such a 200-cycle.
The graph cycle rank is 4001, which gives the stated multiplicities.
Thus $b_3=b_4=1$ and $\Lambda(0)=200$.

\begin{figure}[t]
\centering
\includegraphics[width=0.62\textwidth]{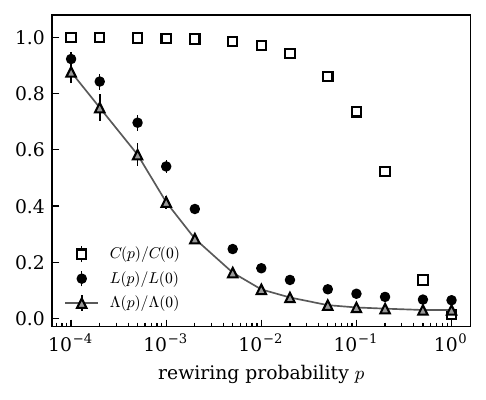}
\caption{\textbf{Normalised statistics in the Watts--Strogatz experiment.} Each point is a mean over twenty connected realisations with $n=1000$ and initial degree $k_0=10$; bars are standard errors. $C$ is average local clustering, $L$ mean shortest-path length and $\Lambda$ the largest barcode death. All are divided by their deterministic ring-lattice values. These curves are computed from the supplied simulations, not reproduced data from the original Watts--Strogatz figure.}
\label{fig:ws-horizon}
\alttext{Three normalised quantities plotted against rewiring probability $p$ on a logarithmic axis from $10^{-4}$ to $1$, with values between $0$ and $1$. The clustering ratio $C(p)/C(0)$, shown by open squares, stays above $0.9$ up to about $p=0.02$ and then falls to near zero at $p=1$. The path-length ratio $L(p)/L(0)$, shown by filled circles, and the horizon ratio $\Lambda(p)/\Lambda(0)$, shown by triangles joined by a line, both start near $0.9$ and decrease; the horizon ratio lies below the path-length ratio.}
\end{figure}

\paragraph{Horizon and path length.}
Let $C(p)$ be the average local clustering coefficient and $L(p)$ the
mean shortest-path length. Figure~\ref{fig:ws-horizon} compares their
sample means with the largest death scale. All quantities are normalised
by their deterministic values at $p=0$. In these samples the normalised
horizon lies below the normalised mean path length. At $p=10^{-3}$,
$L(p)/L(0)$ is about $0.54$ and $\Lambda(p)/\Lambda(0)$ about $0.41$,
and at $p=10^{-2}$ they are about $0.18$ and $0.10$. The reduction in long
cycle relations is already substantial while clustering remains high.
This is an empirical comparison of normalised statistics.
Equation~\eqref{eq:horizon} bounds the horizon by a diameter-based
quantity but does not imply this comparison with mean path length.

\begin{figure}[t]
\centering
\includegraphics[width=0.9\textwidth]{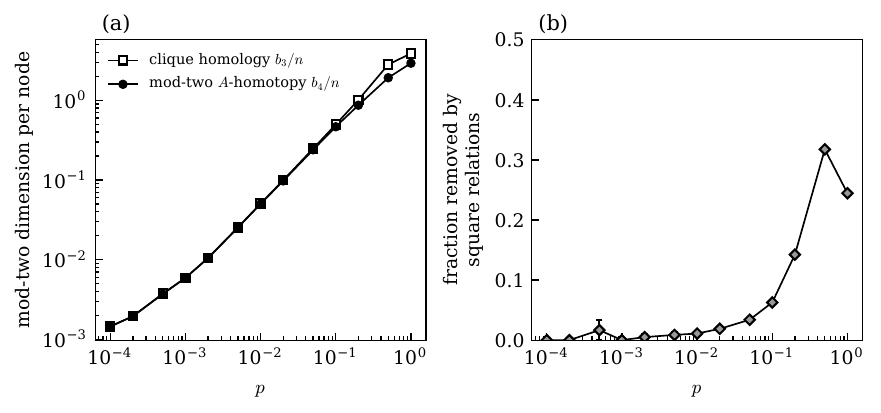}
\caption{\textbf{Mod-two ranks after triangle and square relations.} (a) Means of $b_3/n$ and $b_4/n$. (b) The mean of $1-b_4/b_3$ across the twenty realisations, with standard errors. Every sampled graph has $b_3>0$. This mean fraction differs from a ratio of pooled counts. Here $b_4$ is the dimension of the mod-two abelianisation, not a rank assigned to the full non-abelian group.}
\label{fig:ws-holes}
\alttext{Panel (a) plots, on logarithmic axes, the mean clique-homology dimension $b_3/n$ (open squares) and the mean mod-two $A$-homotopy dimension $b_4/n$ (filled circles) against rewiring probability $p$. Both rise from about $10^{-3}$ at $p=10^{-4}$ to values above one at $p=1$, and the two curves nearly coincide, separating slightly at the largest $p$. Panel (b) plots the mean fraction removed by square relations against $p$. It stays near zero up to about $p=0.01$, rises to a peak of about $0.32$ at $p=0.5$ and is about $0.25$ at $p=1$.}
\end{figure}

\paragraph{Survival after square relations.}
For the sampled probabilities $p\le0.05$, fewer than $4\%$ of the
clique-homology classes are removed at $k=4$, when counts are pooled over
realisations. The two dimension curves in Fig.~\ref{fig:ws-holes}a are
therefore close. The fraction is not identically zero. For example,
at $p=0.05$ the mean ranks are $b_3=247.25$ and $b_4=238.85$, giving a
pooled resolvable fraction of about $3.40\%$. Figure~\ref{fig:ws-holes}b
instead plots the mean of the within-realisation fractions, with standard
errors. The two summaries weight realisations differently.

At larger rewiring probabilities, square relations remove a greater
share of the clique-homology dimension. In the sampled ensemble, this
share is approximately one third near $p=0.5$ and one quarter at $p=1$.
Rewiring preserves the graph cycle rank here, since it preserves the
numbers of vertices and edges. Changes in $b_3$ and $b_4$ reflect changes
in the span of short-cycle relations. They do not establish a one-to-one
correspondence between individual rewired edges and new persistent classes.

\begin{figure}[t]
\centering
\includegraphics[width=0.6\textwidth]{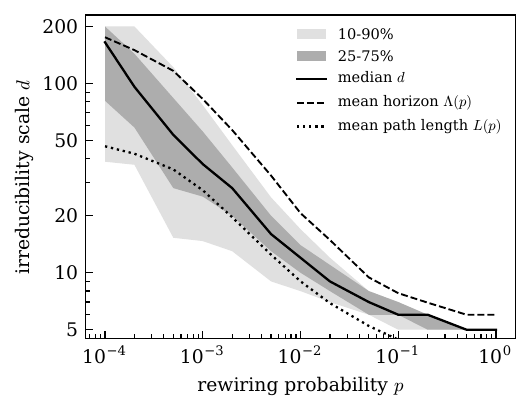}
\caption{\textbf{Death scales above four in the Watts--Strogatz experiment.} The solid line is the median of the pooled deaths, with 25--75\% and 10--90\% ranges shaded. These are distributional ranges, not confidence intervals; realisations with more bars contribute more entries. Dashed and dotted lines show the sample mean horizon and mean path length. The vertical axis measures the death length for the spectrum and the corresponding edge-count lengths for the comparison curves.}
\label{fig:ws-spectrum}
\alttext{Death scale $d$ on a logarithmic axis from $5$ to about $200$, plotted against rewiring probability $p$ from $10^{-4}$ to $1$. Light and dark grey bands show the 10--90\% and 25--75\% ranges of the pooled deaths, and a solid line shows their median. A dashed line shows the mean horizon $\Lambda(p)$ and a dotted line the mean path length $L(p)$. All curves decrease with $p$. The median falls from more than $100$ at the smallest $p$ to about $5$ at $p=1$, and the dotted path-length curve starts near $50$ and stays below the dashed horizon curve.}
\end{figure}

\paragraph{Distribution of death scales.}
Figure~\ref{fig:ws-spectrum} pools the deaths $d>4$ at each probability.
For $10^{-3}\le p\le0.05$, their median is approximately $1.3$--$1.4$
times the sample mean path length, with observed ratios from $1.284$ to
$1.425$. This association is consistent with the role of paths in forming
cycles, but does not provide a unique geometric representative for a bar.
At $p=1$, about $99.6\%$ of the pooled deaths above four occur at five.
Most surviving mod-two classes then disappear at the next allowed cycle
length. The $p=1$ ensemble is the fully rewired model used here, not an
Erd\H{o}s--R\'enyi ensemble.

\subsection{Benchmark networks}
\label{sec:benchmarks}

\begin{figure}[!htbp]
\centering
\includegraphics[width=\textwidth]{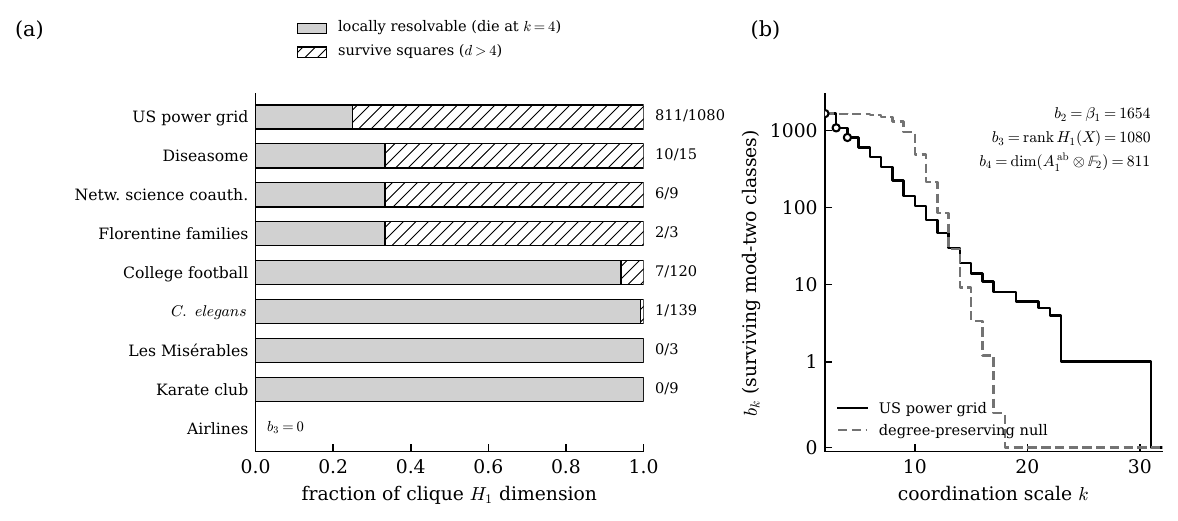}
\caption{\textbf{Mod-two spectra across the benchmarks.} (a) Fractions of clique first-homology dimension removed at $k=4$ and surviving after that stage. Labels give $b_4/b_3$; the airline graph has $b_3=0$ and no fraction is defined. (b) The power-grid survival curve $b_k$ and the mean of five degree-preserving null curves. The vertical scale is linear from zero to one and logarithmic above one, so zeros are shown explicitly. The marked values are $\beta_1$, clique first-homology dimension and mod-two abelianised $A$-dimension.}
\label{fig:atlas}
\alttext{Panel (a) is a horizontal stacked bar chart of the fraction of clique first-homology dimension for nine networks. A grey segment marks classes that die at $k=4$ and a hatched segment classes that survive the square relations. The labels $b_4/b_3$ are $811/1080$ for the US power grid, $10/15$ for the diseasome, $6/9$ for network-science coauthorship, $2/3$ for the Florentine families, $7/120$ for college football, $1/139$ for \emph{C.\ elegans}, $0/3$ for Les Mis\'erables and $0/9$ for the karate club; the airline graph is marked $b_3=0$. Panel (b) is a step plot of the number of surviving mod-two classes $b_k$ against coordination scale $k$ from $2$ to about $32$. The solid curve for the US power grid starts at $b_2=1654$, passes $b_3=1080$ and $b_4=811$, and keeps one class until $k=31$. The dashed curve for the degree-preserving null falls to zero at about $k=18$.}
\end{figure}

Table~\ref{tab:atlas} gives the spectra of the networks of
Table~\ref{tab:data}, and Fig.~\ref{fig:atlas}a shows
how much clique-homology dimension is removed by square relations.
The death multiplicities count independent relations. They are not the
total numbers of triangles or quadrilaterals present in the graph.

\begin{table}[!htbp]
\centering
\small
\caption{\textbf{Mod-two resolvability spectra.} The columns $d=3$ and $d=4$ count barcode deaths, not all short cycles. The dimensions $b_3$ and $b_4$ follow triangle and square filling. In the death column, an exponent is a multiplicity. The final column marks exact triviality established by spanning-tree propagation.}
\label{tab:atlas}
\medskip
\footnotesize
\setlength{\tabcolsep}{4pt}
\begin{tabular}{@{}lrrrrrlrc@{}}
\toprule
Network & $\beta_1$ & $d=3$ & $b_3$ & $d=4$ & $b_4$ & deaths above four & $\Lambda$ & $A_1=1$\\
\midrule
Karate club & 45 & 36 & 9 & 9 & 0 & --- & 4 & yes\\
Les Misérables & 178 & 175 & 3 & 3 & 0 & --- & 4 & yes\\
Airlines & 1063 & 1063 & 0 & 0 & 0 & --- & 3 & yes\\
\emph{C.\ elegans} & 1852 & 1713 & 139 & 138 & 1 & $5$ & 5 & \\
College football & 499 & 379 & 120 & 113 & 7 & $5^6\,6$ & 6 & \\
Florentine families & 6 & 3 & 3 & 1 & 2 & $5^2$ & 5 & \\
Network science & 536 & 527 & 9 & 3 & 6 & $5^2\,7^3\,8$ & 8 & \\
Diseasome & 1320 & 1305 & 15 & 5 & 10 & $5^5\,6^2\,7^2\,9$ & 9 & \\
US power grid & 1654 & 574 & 1080 & 269 & 811 & $5$--$31$ (18 scales) & 31 & \\
\bottomrule
\end{tabular}
\end{table}

\paragraph{Complete resolution at length four.}
The karate and Les Mis\'erables graphs have $b_3=9$ and $b_3=3$,
respectively, but $b_4=0$. For both, the spanning-tree propagation test
also certifies $A_1=1$. The airline graph has $b_3=b_4=0$, and the same
test certifies triviality there. Thus in these examples the square
relations suffice to remove all clique-homology classes. This is a result
about the specified moves. It does not imply that the original homology
classes are spurious or that corresponding rearrangements take place in
the social or transport systems.

\paragraph{Small surviving dimensions.}
For \emph{C.\ elegans}, $b_3=139$ and $b_4=1$, and the surviving bar dies
at five. In the football graph, $b_3=120$ and $b_4=7$, with six deaths
at five and one at six. Since these groups are free
(Section~\ref{sec:freecomplex}), the $A$-groups are $\ZZ$ and $F_7$, and
the barcode determines the rest of each sequence.

\paragraph{Longer spectra.}
The coauthorship, diseasome and Florentine networks retain respectively
six of nine, ten of fifteen, and two of three clique-homology dimensions.
Their longest deaths are eight, nine and five. The power grid retains
811 of 1080 dimensions, with deaths at eighteen distinct integer lengths
between five and 31. Networks of similar surviving dimension can therefore
still differ in the length of relations needed to remove those classes.

\paragraph{Comparison with degree-preserving rewiring.}
The saved power-grid comparison uses five randomisations obtained by
connected degree-preserving edge swaps, each requesting ten times the
number of edges in swap attempts. The seeds and procedure are supplied.
The five spectra retain 1637--1644 of 1654 dimensions at $k=4$, but their
longest deaths are 16--18. The real graph has many more independent short
relations, with 574 deaths at three and 269 at four, yet a longer tail.
The real survival curve crosses the mean null curve near $k=13$.

This comparison shows that the recorded spectrum is not reproduced by
these particular degree-preserving randomisations. It is consistent with
an influence of spatial constraints, but it does not isolate that
influence. Rewiring also changes clustering and community structure, and
five realisations provide only an illustrative null comparison. A study
of spatial mechanisms would require a larger ensemble, checks of the swap
procedure and a null model retaining suitable geometric constraints.

\paragraph{The complex level.}
Passing from $G$ to $\Gmax{1}(X(G))$ changes the object being analysed.
For the karate and Les Mis\'erables networks, the $A$-groups of all its
components are certified trivial. Its aggregate mod-two $A$-rank is
three for football and one for each of the diseasome and power-grid
networks. These need not agree with $b_4(G)$, because loops of overlapping
maximal cliques differ from loops of network vertices. The groups of
all components are free (Section~\ref{sec:freecomplex}).

\subsection{Loops of cliques around a hub}
\label{sec:ego}

Every ego clique complex is contractible, but the maximal-clique
connectivity graph need not have trivial $A$-groups. We enumerated all
6980 egos in the karate, Les Mis\'erables, Florentine, football,
coauthorship, diseasome and power-grid networks. Three have a positive
mod-two $A$-rank. In each case an edge cochain certifies non-triviality.
For all other 6977 egos, the spanning-forest propagation certificate
proves triviality in every component. Thus the classification within
these seven data graphs concerns the full groups, not just a zero-rank
test. It is not an estimate of their frequency in networks generally.

\begin{figure}[!htbp]
\centering
\includegraphics[width=\textwidth]{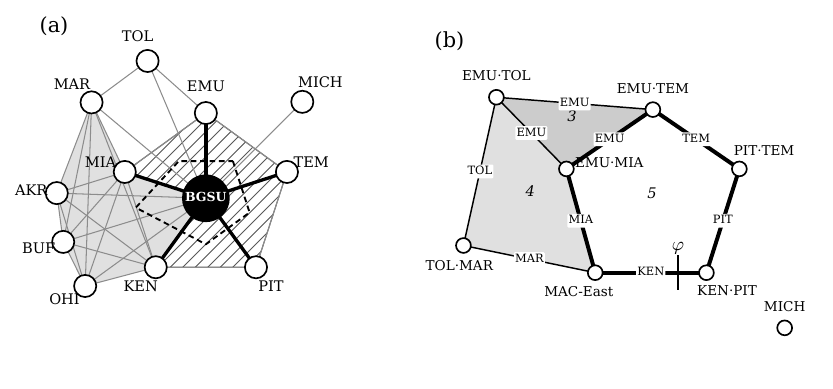}
\caption{\textbf{A loop of cliques around Bowling Green State.} (a) Edges mean games in the 2000 football dataset. Four hatched triangles and the larger shaded MAC-East clique form the displayed ring of overlapping maximal cliques. (b) Their full connectivity graph has two components, including the isolated MICH vertex. In the seven-vertex component, filling the triangle and square leaves fundamental group $\ZZ$. The marked edge belongs to no triangle or square, and its cochain detects the pentagon. The ego clique complex in (a) has zero reduced homology. The exact group claim in (b) concerns the non-trivial component, with a basepoint there.}
\label{fig:bgsu}
\alttext{Panel (a) shows the ego network of Bowling Green State, drawn as a black central vertex, with its eleven neighbours labelled by abbreviated team names. A grey region marks the MAC-East clique, four triangles through the centre are hatched, and a dashed pentagon around the centre indicates the loop of cliques. Panel (b) shows the connectivity graph of the maximal cliques. Seven vertices, labelled by clique members, form one component containing a shaded triangle, a shaded quadrilateral and a pentagon drawn with heavy lines. An eighth vertex, MICH, is isolated. Edges are labelled by the shared opponent, and the edge through KEN is marked $\varphi$.}
\end{figure}

\paragraph{Bowling Green State.}
The ego in Fig.~\ref{fig:bgsu} has eleven neighbours. Its connectivity
graph has eight vertices, nine edges and two components. One component
is the isolated clique consisting of Bowling Green and Michigan. The
other has a triangle, a chordless square and a pentagon as a cycle basis.
Removing the triangle through the EMU--TEM to EMU--TOL edge, then the
square through the EMU--MIA to EMU--TOL edge, leaves a connected graph
of cycle rank one. Lemma~\ref{lem:cert}(c) proves that its $A$-group is
$\ZZ$. The isolated component has trivial group.

The displayed pentagon passes through four maximal triangles and the
larger MAC-East clique. Consecutive cliques share Bowling Green and one
opponent. The edge marked $\varphi$ in the figure belongs to no triangle
or square in the connectivity graph. Its cochain evaluates to one on
the pentagon, providing a second, independent type of certificate for
non-triviality.

\paragraph{Tennessee.}
The Tennessee ego gives a connected clique-connectivity graph with ten
vertices, twenty edges and cycle rank eleven. It contains thirteen
triangles and 27 distinct squares. A five-clique loop passes through
shared opponents Southern Mississippi, Alabama, Arkansas,
Louisiana--Monroe and Memphis. The connectivity edge shared through
Southern Mississippi belongs to no triangle or square and certifies a
non-trivial loop. The mod-two $A$-rank is one. The freeness test of
Section~\ref{sec:freecomplex} identifies its $A$-group as $\ZZ$.

\paragraph{Colon cancer.}
The colon-cancer ego in the mixed diseasome file has 49 maximal cliques.
Its connectivity graph has 134 edges and seventeen components, so its
aggregate cycle rank is $134-49+17=102$. Its 287 triangles and 1840
squares leave a one-dimensional mod-two quotient at $k=4$. A witness
loop runs through cliques shared via renal cell carcinoma,
von Hippel--Lindau syndrome, leukaemia, breast cancer and hepatic adenoma.
The edge shared through von Hippel--Lindau syndrome lies on no short
cycle. The witness uses disease vertices, but its certificate is checked
in the connectivity graph of the full mixed ego network. The freeness test of
Section~\ref{sec:freecomplex} identifies its group as $\ZZ$.

These examples compare the original ego complex with a graph constructed
from its maximal cliques. Their non-triviality is visible to homology
of the corresponding triangle-and-square cell complex, even though
homology of the original ego complex vanishes in positive degrees.

\section{Discussion}
\label{sec:discussion}

Persistent $A$-homotopy makes the permitted deformation of a loop an
explicit part of a network model. For a scientist working with such a
model, two separate questions then arise: which interactions are present,
and which combinations of interactions count as a local rearrangement of
a loop. A threshold filtration addresses the first question. The coordination filtration
addresses the second while keeping the graph fixed. Their combination
allows a class to be long-lived with respect to the network threshold but
short-lived with respect to the relations one is willing to impose.

For the networks examined in Section~\ref{sec:free}, the barcode carries
all the information in the group-valued construction. Every group in
their sequences was certified free, so by
Proposition~\ref{prop:freetower} the barcode determines the groups and the
maps between them. Section~\ref{sec:whyfree} suggests why. In sparse
networks the short cycles rarely overlap, and in clustered networks many
short cycles lie in neighbourhoods that the deletion of dominated vertices
removes. These are explanations and not a proof for any given network.
Freeness has to be checked case by case, and the test of
Section~\ref{sec:freetest} does this for networks with thousands of
vertices.

The group records more than the barcode in the periodic examples of
Section~\ref{sec:periodic}. On a torus the two basis loops commute. In
the random geometric graphs on the torus, the group changed from $F_2$ to
$\ZZ^2$ in six of nine samples at a length where no bar ends. In a network
repeated over periodic time, each loop of the network commutes with the
loop around the period, and the barcode cannot distinguish this from an
additional independent cycle. Babson's results~\cite{Babson2012} give a different,
asymptotic case, in which the group at length three is non-trivial while $b_3=0$. Our
random graphs in that range of edge probabilities, with at most 1500
vertices, had not reached it. Spatial models with periodic boundary
conditions, lattice models and temporal networks observed over a periodic
schedule have the two independent periodic directions of
Section~\ref{sec:periodic}. In such applications we would compute the
groups, or at least run the freeness test, before reading the barcode as a
complete description.

The mod-two spectrum provides an accessible entry point. Its relation to
minimum cycle bases makes it computable with existing methods, and its
successive rank drops quantify the effect of short-cycle relations.
This is also where the connection to persistent homology is exact.
The spectrum is the first homology barcode of $Y_k(G)$, and the $k=3$
row of the two-parameter construction is clique first homology persistence.
Thus persistent homology supplies the linear tools, while the
$A$-homotopy construction specifies a sequence of loop equivalences and
retains a group-level object beyond that summary. The two viewpoints
are complementary. Neither is a universal replacement for the other,
and the available information depends on the complex and the relations
chosen before an invariant is computed.

The examples show what can be learned without assigning a physical
mechanism to every class. On square lattices, the choice to fill
plaquettes reproduces the fundamental group of the intended surface.
The Watts--Strogatz results show how the distribution of required cycle
lengths changes with rewiring, alongside clustering and path length.
The power-grid spectrum differs from the five degree-preserving null
spectra in both its short relations and its long tail. That result
motivates a comparison with spatially constrained models, but it does not
establish spatial embedding as the sole explanation.

The ego examples concern a further level of organisation. A hub makes
its ego clique complex contractible, but it need not remove loops among
overlapping maximal cliques. In the Bowling Green, Tennessee and colon-cancer
examples the clique-connectivity graph has a component with group $\ZZ$.
These are statements about overlap
patterns. They can be useful when cliques represent interacting groups,
but they do not by themselves establish a coordination bottleneck,
causal dependence or dynamical instability in the original system.

Several limitations guide the next steps. The freeness test decided every
case reported here, but it is a procedure that may leave a presentation
undecided, not an algorithm for all presentations. For the periodic grids
with shortcuts, it proved that most of the groups are not free without
identifying them. For the Watts--Strogatz
experiment, freeness rests on seven graphs generated with its parameters,
not on its 260 saved realisations. Mod-two abelianisation does not detect odd
torsion in the abelianisation, and its dimension bounds the minimum number
of group generators from below. Maximal-clique enumeration can be
expensive, and the ego scan here excludes the two densest benchmarks.
The null comparison is small, and the mixed diseasome and undocumented
airline sample call for caution in domain-specific interpretation.
The cycle-length parameter also counts edges rather than physical distance
or the cost of a rearrangement.

For weighted or spatial networks, an application should first identify
which local moves are admissible and how their cost is measured. A
filtration by those costs could then be compared with the graph-threshold
filtration, with its definitions and stability properties proved for that
setting. A non-nested temporal sequence requires its own treatment.
For functional brain networks, for example, correlations alone do not
justify interpreting a contractible loop as a feasible neural operation.
Such an interpretation would require a model connecting the allowed moves
to dynamics or interventions.

Stability under edge insertions is proved in the companion
paper~\cite{Rajkovic2026b}. A separate treatment could study functoriality
under broader classes of maps. Two questions raised here remain open.
We do not know whether the groups of every planar graph are free, and we
have not studied networks over periodic time that change during the
period, whose graphs are not Cartesian products. The present paper
has a more focused aim. It makes loop relations visible as a modelling
choice, identifies when the mod-two barcode describes the whole
construction and gives network researchers a way to compare that choice
with the threshold persistence they already use.

\section*{Data and code availability}

The accompanying reproducibility package contains the input networks,
retrieval locations and checksums, simplified edge lists, the 260 saved
Watts--Strogatz records and the five saved power-grid null spectra.
Python scripts recompute the benchmark spectra, enumerate the seven-network
ego scan, export and check certificates, and regenerate the vector figures.
MATLAB reference routines implement the core calculations and include
self-tests and example scripts. Python scripts for the freeness test of
Section~\ref{sec:freetest} generate the model graphs of
Tables~\ref{tab:free} and~\ref{tab:torus} with fixed seeds, contract the
street networks, run the eliminations and the tests for non-freeness, and
log the outcome at every length. The scripts and logs are included.

\section*{Use of AI tools}
The author used the AI systems Claude (Anthropic) and Codex (OpenAI) in preparing this paper. These tools were used for editorial assistance, verification of network computations and checking of mathematical arguments and citations. 

\section*{Conflict of Interest}
None declared.

\section*{Funding}
No funding was received to assist with the preparation of this manuscript.


\begin{thebibliography}{99}
\raggedright
\bibitem{Battiston2020} F.~Battiston, G.~Cencetti, I.~Iacopini, V.~Latora,
M.~Lucas, A.~Patania, J.-G.~Young and G.~Petri, Networks beyond pairwise
interactions: structure and dynamics, \emph{Phys.\ Rep.} \textbf{874}, 1--92
(2020).
\bibitem{HMR2009} D.~Horak, S.~Maleti\'c and M.~Rajkovi\'c, Persistent
homology of complex networks, \emph{J.\ Stat.\ Mech.} P03034 (2009).
\bibitem{Petri2013} G.~Petri, M.~Scolamiero, I.~Donato and F.~Vaccarino,
Topological strata of weighted complex networks, \emph{PLoS ONE}
\textbf{8}, e66506 (2013).
\bibitem{Atkin1974} R.~H.~Atkin, \emph{Mathematical Structure in Human
Affairs} (Heinemann, London, 1974).
\bibitem{Atkin1977} R.~H.~Atkin, \emph{Combinatorial Connectivities in Social
Systems} (Birkh\"auser, Basel, 1977).
\bibitem{BKLW2001} H.~Barcelo, X.~Kramer, R.~Laubenbacher and C.~Weaver,
Foundations of a connectivity theory for simplicial complexes, \emph{Adv.\
Appl.\ Math.} \textbf{26}, 97--128 (2001).
\bibitem{BL2005} H.~Barcelo and R.~Laubenbacher, Perspectives on $A$-homotopy
theory and its applications, \emph{Discrete Math.} \textbf{298}, 39--61 (2005).
\bibitem{BBdLL2006} E.~Babson, H.~Barcelo, M.~de~Longueville and
R.~Laubenbacher, Homotopy theory of graphs, \emph{J.\ Algebraic Combin.}
\textbf{24}, 31--44 (2006).
\bibitem{Rajkovic2026a} M.~Rajkovi\'c, $A$-homotopy theory for $Q$-analysis
and complex networks: an expository introduction, manuscript (2026).
\bibitem{MemoliZhou} F.~M\'emoli and L.~Zhou, Persistent homotopy groups of
metric spaces, \emph{J.\ Topol.\ Anal.} \textbf{17}, 1481--1542 (2025).
See also \url{https://arxiv.org/abs/1912.12399}.
\bibitem{KapulkinKershaw} K.~Kapulkin and N.~Kershaw, Topological data analysis
using persistent discrete homology, \url{https://arxiv.org/abs/2506.15020}.
\bibitem{Rajkovic2026b} M.~Rajkovi\'c, Toward a filtration of local
resolvability in $A$-homotopy theory, manuscript (2026).
\bibitem{Babson2012} E.~Babson, Fundamental groups of random clique
 complexes, \url{https://arxiv.org/abs/1207.5028} (2012).
\bibitem{BS2008} H.~Barcelo and S.~Smith, The discrete fundamental group of
 the order complex of $B_n$, \emph{J.\ Algebraic Combin.} \textbf{27},
 399--421 (2008). See also \url{https://arxiv.org/abs/0711.0915}.
\bibitem{Davis1941} A.~Davis, B.~B.~Gardner and M.~R.~Gardner, \emph{Deep
South} (University of Chicago Press, Chicago, 1941).
\bibitem{NXdata} {[dataset]} NetworkX developers, graph data distributed
 with NetworkX 3.7: \texttt{karate\_club\_graph},
 \texttt{les\_miserables\_graph}, \texttt{florentine\_families\_graph}
 and \texttt{davis\_southern\_women\_graph}, \url{https://networkx.org}.
\bibitem{Horton1987} J.~D.~Horton, A polynomial-time algorithm to find the
shortest cycle basis of a graph, \emph{SIAM J.\ Comput.} \textbf{16},
358--366 (1987).
\bibitem{Kavitha2009} T.~Kavitha, C.~Liebchen, K.~Mehlhorn, D.~Michail,
R.~Rizzi, T.~Ueckerdt and K.~A.~Zweig, Cycle bases in graphs:
characterization, algorithms, complexity, and applications, \emph{Comput.\
Sci.\ Rev.} \textbf{3}, 199--243 (2009).
\bibitem{Padgett1993} J.~F.~Padgett and C.~K.~Ansell, Robust action and the
rise of the Medici, 1400--1434, \emph{Am.\ J.\ Sociol.} \textbf{98},
1259--1319 (1993).
\bibitem{LS1977} R.~C.~Lyndon and P.~E.~Schupp, \emph{Combinatorial Group
 Theory} (Springer, Berlin, 1977).
\bibitem{igraph} G.~Cs\'ardi and T.~Nepusz, The igraph software package for
complex network research, \emph{InterJournal Complex Systems} 1695 (2006).
\bibitem{networkx} A.~A.~Hagberg, D.~A.~Schult and P.~J.~Swart, Exploring
network structure, dynamics, and function using NetworkX, in
\emph{Proceedings of the 7th Python in Science Conference} (2008), pp.~11--15.
\bibitem{Hatcher2002} A.~Hatcher, \emph{Algebraic Topology} (Cambridge
 University Press, Cambridge, 2002).
\bibitem{NewmanData} {[dataset]} M.~E.~J.~Newman, network data collection,
 college football network, file \texttt{football.gml},
 \url{https://websites.umich.edu/~mejn/netdata/football.zip},
 accessed 11 October 2026.
\bibitem{GephiData} {[dataset]} Gephi, Sample datasets: files
 \texttt{netscience.gml}, \texttt{celegans.gexf}, \texttt{diseasome.gexf},
 \texttt{airlines.graphml} and \texttt{power.gml},
 \url{https://docs.gephi.org/desktop/User_Manual/Datasets/},
 accessed 11 October 2026. Archive URLs and SHA-256 checksums are given
 in the supplementary data manifest.
\bibitem{Goh2007} K.-I.~Goh, M.~E.~Cusick, D.~Valle, B.~Childs, M.~Vidal and
A.-L.~Barab\'asi, The human disease network, \emph{Proc.\ Natl.\ Acad.\ Sci.\
USA} \textbf{104}, 8685--8690 (2007).
\bibitem{Zachary1977} W.~W.~Zachary, An information flow model for conflict
and fission in small groups, \emph{J.\ Anthropol.\ Res.} \textbf{33},
452--473 (1977).
\bibitem{Knuth1993} D.~E.~Knuth, \emph{The Stanford GraphBase: A Platform for
Combinatorial Computing} (Addison-Wesley, Reading, MA, 1993).
\bibitem{GN2002} M.~Girvan and M.~E.~J.~Newman, Community structure in social
and biological networks, \emph{Proc.\ Natl.\ Acad.\ Sci.\ USA} \textbf{99},
7821--7826 (2002).
\bibitem{Newman2006} M.~E.~J.~Newman, Finding community structure in networks
using the eigenvectors of matrices, \emph{Phys.\ Rev.\ E} \textbf{74}, 036104
(2006).
\bibitem{White1986} J.~G.~White, E.~Southgate, J.~N.~Thomson and
S.~Brenner, The structure of the nervous system of the nematode
\emph{Caenorhabditis elegans}, \emph{Phil.\ Trans.\ R.\ Soc.\ Lond.\ B}
\textbf{314}, 1--340 (1986).
\bibitem{WS1998} D.~J.~Watts and S.~H.~Strogatz, Collective dynamics of
`small-world' networks, \emph{Nature} \textbf{393}, 440--442 (1998).
\bibitem{BA1999} A.-L.~Barab\'asi and R.~Albert, Emergence of scaling in
 random networks, \emph{Science} \textbf{286}, 509--512 (1999).
\bibitem{HolmeKim2002} P.~Holme and B.~J.~Kim, Growing scale-free networks
 with tunable clustering, \emph{Phys.\ Rev.\ E} \textbf{65}, 026107 (2002).
\bibitem{MolloyReed1995} M.~Molloy and B.~Reed, A critical point for random
 graphs with a given degree sequence, \emph{Random Struct.\ Algorithms}
 \textbf{6}, 161--179 (1995).
\bibitem{Krioukov2010} D.~Krioukov, F.~Papadopoulos, M.~Kitsak, A.~Vahdat and
 M.~Bogu\~n\'a, Hyperbolic geometry of complex networks, \emph{Phys.\ Rev.\ E}
 \textbf{82}, 036106 (2010).
\bibitem{LFR2008} A.~Lancichinetti, S.~Fortunato and F.~Radicchi, Benchmark
 graphs for testing community detection algorithms, \emph{Phys.\ Rev.\ E}
 \textbf{78}, 046110 (2008).
\bibitem{Holland1983} P.~W.~Holland, K.~B.~Laskey and S.~Leinhardt,
 Stochastic blockmodels: first steps, \emph{Social Networks} \textbf{5},
 109--137 (1983).
\bibitem{Ispolatov2005} I.~Ispolatov, P.~L.~Krapivsky and A.~Yuryev,
 Duplication-divergence model of protein interaction network, \emph{Phys.\
 Rev.\ E} \textbf{71}, 061911 (2005).
\bibitem{Penrose2003} M.~Penrose, \emph{Random Geometric Graphs} (Oxford
 University Press, Oxford, 2003).
\bibitem{ErdosRenyi1960} P.~Erd\H{o}s and A.~R\'enyi, On the evolution of
 random graphs, \emph{Publ.\ Math.\ Inst.\ Hung.\ Acad.\ Sci.} \textbf{5},
 17--61 (1960).
\bibitem{OSMHelsinki} {[dataset]} OpenStreetMap contributors, extracts
 \texttt{Helsinki.osm.pbf} and \texttt{test.osm.pbf} distributed with the
 Python package \texttt{pyrosm} 0.29.0, \url{https://pypi.org/project/pyrosm/}.
 OpenStreetMap data are available under the Open Database Licence.
\bibitem{OSMOakland} {[dataset]} OpenStreetMap contributors, extract
 \texttt{West-Oakland.osm.bz2} in the test data of the \texttt{OSMnx}
 repository, \url{https://github.com/gboeing/osmnx}, accessed
 7 November 2026.
\bibitem{CFH2015} A.~Costa, M.~Farber and D.~Horak, Fundamental groups of
 clique complexes of random graphs, \url{https://arxiv.org/abs/1312.1208}.
\bibitem{Barthelemy2011} M.~Barth\'elemy, Spatial networks, \emph{Phys.\
 Rep.} \textbf{499}, 1--101 (2011).
\bibitem{HolmeSaramaki2012} P.~Holme and J.~Saram\"aki, Temporal networks,
 \emph{Phys.\ Rep.} \textbf{519}, 97--125 (2012).
\bibitem{Adamaszek2013} M.~Adamaszek, Clique complexes and graph powers,
 \emph{Israel J.\ Math.} \textbf{196}, 295--319 (2013).
 \url{https://arxiv.org/abs/1104.0433}.
\end{thebibliography}
\end{document}